\documentclass{svmult}
\usepackage{makeidx}         
\usepackage{graphicx}        
\usepackage{multicol}        
\usepackage[bottom]{footmisc}

\makeindex             

\usepackage{epsfig}
\usepackage{graphics}
\usepackage{dcolumn}
\usepackage{bm}

\def\lnsco{La$_{2-x-y}$Nd$_{y}$Sr$_{x}$CuO$_{4}$\,}
\def\lsco{La$_{2-x}$Sr$_{x}$CuO$_{4}$\,}
\def\lco{La$_{2}$CuO$_{4}$\,}
\def\ls18{La$_{1.875}$Sr$_{0.125}$CuO$_{4}$\,}
\def\lns0418{La$_{1.475}$Nd$_{0.4}$Sr$_{0.12}$CuO$_{4}$\,}

\def\scocl{Sr$_{2}$CuO$_{2}$Cl$_{2}$\,}
\def\ybco{YBa$_{2}$Cu$_{3}$O$_{6 + \delta}$\,}
\def\lbco{La$_{2-x}$Ba$_{x}$CuO$_{4}$\,}
\def\cm-1{cm$^{-1}$}

\begin{document}

\title*{Magnetic and Charge Correlations  in \lnsco: Raman Scattering Study}
\author{A. Gozar\inst{1,2}, Seiki Komiya\inst{3}, Yoichi Ando\inst{3} 
\and G. Blumberg\inst{1,*}}
\institute{
Bell Laboratories, Lucent Technologies, Murray Hill, NJ 07974, USA \and 
University of Illinois at Urbana-Champaign, Urbana, IL 61801, USA \and
Central Research Institute of Electric Power Industry, Komae, Tokyo 
201-8511, Japan
\texttt{
\begin{center}
    (\textit{Frontiers in Magnetic Materials} (Ed. A.V. Narlikar), 
    Spinger-Verlag 2005, pp. 755-789).
\end{center}
}} 

\maketitle

\section{The Phase Diagram and Structural Properties of the High Temperature Superconductor \lsco}

\lsco is one of the most studied Cu-O based layered perovskites \cite{KastnerRMP98}.
It exhibits some of the most important aspects related to the physics of strongly correlated electrons and, more important, is one of the compounds which belong to the family of high temperature superconducting cuprates.
In fact the high T$_{c}$ superconductivity (SC) rush which began in 1986 started with a variant of \lsco, a Ba-La-Cu-O based compound \cite{BednorzZPB86}, where the authors observed a highest onset SC temperature T$_{c}$ in the 30~K range.

\begin{figure}[t]
\centerline{
\epsfig{figure=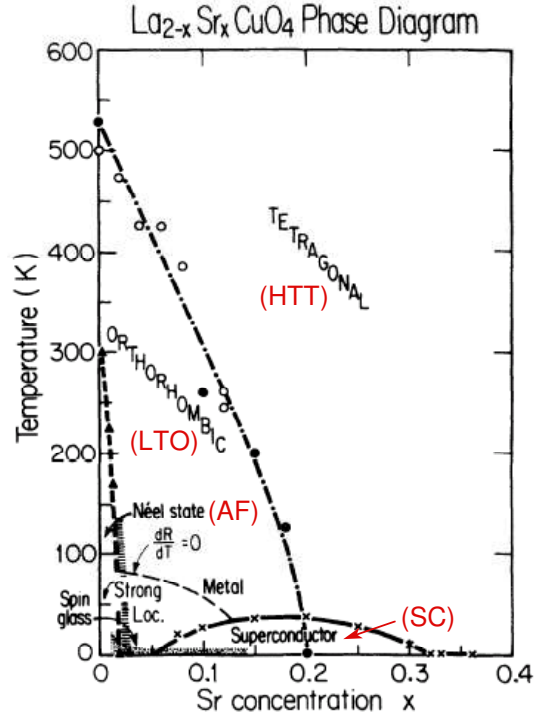,width=70mm}
}
\caption{
Phase diagram of \lsco from Ref.~\cite{KeimerPRB92}.
HTT and LTO stand for the high temperature tetragonal and low temperature orthorhombic phases respectively.
AF denotes the long ranged antiferromagnetic region (N\'{e}el state) at low dopings and SC denotes the superconductivity dome with a maximum around $x = 0.2$ Sr concentration. 
}
\label{f11}
\end{figure}

\begin{figure}[t]
\centerline{
\epsfig{figure=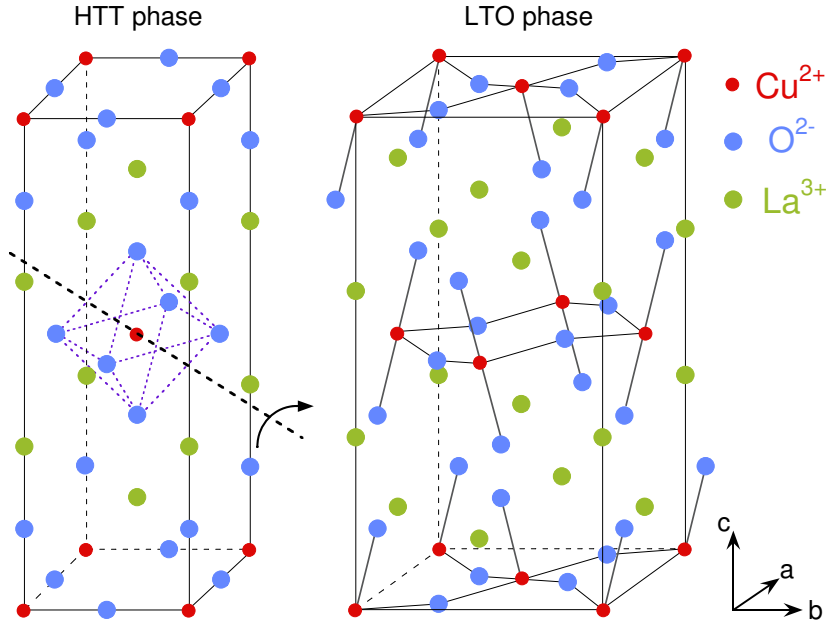,width=110mm}
}
\caption{
The layered perovskite structure of \lsco.
Left: the high temperature tetragonal (HTT) phase where the CuO$_{2}$ planes are flat.
Right: the low temperature orthorhombic (LTO) phase, $(Bmab)$ space group which is obtained from the HTT structure by rigid CuO$_{6}$ octahedra rotations as shown.
In the $Bmab$ setting the $b$-axis is parallel to the corrugation in the CuO$_{2}$ planes.
}
\label{f12}
\end{figure}

The phase diagram of \lsco is shown in Fig.~\ref{f11}, see also Ref.~\cite{KeimerPRB92}.
Several electronic ground states as well as structural phases evolve with Sr concentration.
For $x$(Sr) $\leq 0.02$ the crystals have long range antiferromagnetic (AF) order and one can observe a very rapid suppression of the N\'{e}el ordering temperature T$_{N}$ with the amount of Sr.
While for $x = 0$ the AF transition is slightly above room temperature, T$_{N}$ decreases in the 150~-~200~K range for $x = 0.01$ and it is completely suppressed above $x = 0.02$.
The phase diagram shows also a SC dome starting at $x = 0.05$ and ending around $x~=~0.32$.
The maximum T$_{c}$ of about 40~K is reached at the optimal doping $x = 0.2$.
The highest SC temperature ($T_{c} = 51.5$~K) in the \lsco family was achieved in thin films under epitaxial strain \cite{BozovicPRL02}.
There are also two structural phases of the this compound.
One is tetragonal and the other one is orthorhombic, see Fig.~\ref{f12}.
Sr substitution for La decreases the orthorhombicity and the crystal remains tetragonal at all temperatures at values of $x$(Sr) which correspond roughly to the region of maximum T$_{c}$.
Other intervening phases shown in Fig.~\ref{f11}, spin glass at low temperatures and low Sr concentration, Fermi or non-Fermi liquid behavior depending on if one is in the far right side of the phase diagram or not, are discussed in literature \cite{KastnerRMP98}.

The crystal structure of \lsco is shown in Fig.~\ref{f12}.
The occurrence of several structural phases is typical for perovskites and they generally happen as a result of the lattice strain between the rare-earth and the CuO$_{2}$ layers.
The strain is often released by various bucklings of the transition metal - oxygen planes and this is also the case here.
The HTT phase has flat CuO$_{2}$ planes and the transition to the LTO phase can be understood within a good approximation as a rigid rotation of the CuO$_{6}$ octahedra around an axis making 45$^{\circ}$ with respect to the orthorhombic axes.
As a result, half of the O atoms will be situated above and the other half below the plane determined by the Cu atoms, see Fig.~\ref{f12}.
The lattice constants of the LTO phase at low temperatures are $a = 5.354$~\AA, \ $b = 5.401$~\AA \ and $c = 13.153$~\AA.
\ So the orthorhombicity, defined by $2(a - b) / (a + b)$, is small, only of about 0.8\%.

In the parent compound, \lco, one has La$^{3+}$ and O$^{2-}$ non-magnetic ions so copper will be in a  Cu$^{2+}$ oxidation state to insure neutrality.
As a result, the last Cu $3d^{9}$ shell will contain a hole carrying a spin $S = 1/2$ which is responsible for the magnetic properties.
Sr$^{2+}$ substitution for La leads to hole doping of the CuO$_{2}$ planes.
It is believed that hole pairing and the acquirement of 3D coherence lead to the occurrence of superconductivity.

In this chapter we will also talk about certain properties of Nd doped \lsco and we mention here some well established effects associated with Nd substitution for La.
One is that Nd in \lnsco suppresses superconducting correlations.
For instance magnetic susceptibility data in \lnsco with $x = 0.2$ show that SC vanishes for values of $y$ greater than about 0.6 \cite{BuchnerPRL94}.
Another effect is that this suppression of SC is accompanied by the enhancement of other types of correlations, the appearance of the so called 'stripes' \cite{TranquadaNature96}, which are proposed to be quasi-1D in plane charge and/or spin super-modulations.
While the discussion above suggests that these two states act against each other, it is not clear at this moment if the stripes are helping or competing with SC.
Another effect is related to changes in the crystal structure as a result of inter-layer chemical modifications.
Nd doping brings in another phase, the low temperature tetragonal (LTT) structure, which can be imagined as a rigid CuO$_{6}$ octahedra tilt around the axis whose vector is defined by $1/\sqrt{2} (\hat{a} + \hat{b})$ where $a$ and $b$ are the orthorhombic axes of the LTO phase.

In the following we will discuss low energy magnetic properties of \lsco at light $x$(Sr) doping level.
Although much is known about the physics of 2D $S = 1/2$ antiferromagnets, there are recent experiments which show surprising properties in macroscopically orthorhombic crystals in the presence of external magnetic fields.
It is worth mentioning in this respect that recent neutron scattering in such crystals studies show that even the crystal structure has not rigorously been determined yet \cite{KeimerPrivate} although the deviations from the $Bmab$ symmetry may be very small.
We will show later in this chapter, especially in connection to the phononic and electronic properties, that the effects of orthorhombicity are surprisingly large.
In the following we discuss long wavelength spin-wave excitations as a function of temperature, doping and magnetic field.
We show that the low energy spin dynamics allows us to observe a spin ordered state induced by magnetic fields, a state which persist up to quite high temperatures in crystals with long range AF order \cite{GozarPRL04}.
It will be shown that although the orthorhombicity is small, there are dramatic anisotropy effects in the in plane electronic and phononic excitations.
Our data indicate that at commensurate hole doping $x = 1/8$ \lnsco and independent of Nd concentration there are local deviations in the crystal structure due to a spread in the CuO$_{6}$ tilt angle.
We will discuss this behavior in connection with possible spin and charge modulations in the CuO$_{2}$ planes \cite{GozarPRB03}.

\section{Magnetic and Electronic Properties of Macroscopically 
Orthorhombic \lsco at Light Doping ($0 \leq x \leq 0.03$)}

\subsection{Why is a Study of Low Energy Magnetism Interesting?}

SC as well as the normal properties of 2D Mott-Hubbard systems have already triggered a lot of effort to understanding the evolution of the ground state and of the AF correlations as a function of doping.
However, in spite of the small orthorhombicity, the impact of the low energy magnetism on the carrier and lattice dynamics in \emph{detwinned} \lsco crystals has recently been shown to be significant and surprising new effects were found.

\begin{figure}[t]
\centerline{
\epsfig{figure=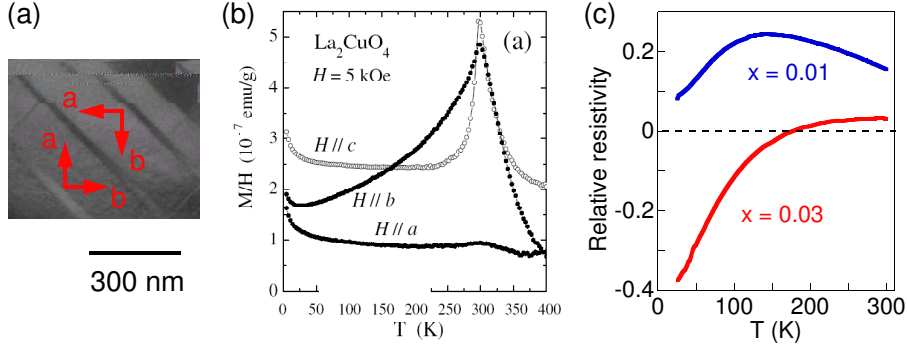,width=120mm}
}
\caption{
(a) Typical orthorhombic domains in a twinned \lsco sample (from Ref.~\cite{HoribePRB00}).
(b) Magnetic susceptibility in a detwinned single crystal of \lco (data from Ref.~\cite{LavrovPRL01}).
(c) Relative resistivity defined by $2(\rho_{a} - \rho_{b}) / (\rho_{a} + \rho_{b})$ in x~=~0.01 and 0.03 detwinned \lsco crystals (see also Ref.~\cite{AndoPRL02}).
}
\label{f13}
\end{figure}

What does detwinned mean in the first place?
On cooling from the HTT to the LTO phase the crystal develops orthorhombic domains, called twins, on the nanometer to micron scale.
The sign of the orthorhombic distortions changes across the twin boundaries, as shown in Fig.~\ref{f13}a.
Accordingly, for a macroscopic probe (and a Raman setup which uses a focussed laser spot larger than about several $\mu$ diameter is an example) the sample looks effectively tetragonal.
If uniaxial pressure of about 15-30 MPa is applied while slowly cooling the crystal through the HTT-LTO phase transition, a detwinned, i.e. macroscopically orthorhombic, crystal can be grown \cite{LavrovPRL01}.

This leads to non-trivial effects if one looks in Fig.~\ref{f13}b-c.
The magnetization data shows two peaks at the N\'{e}el transition of around 300~K in \lco and the magnetic anisotropy is preserved in a wide range of temperatures above T$_{N}$.
The susceptibility along the $a$-axis, $\chi_{a} (T)$, is featureless showing that this axis is magnetically inert, at least at small fields.
The structure with two peaks is due to the various spin anisotropy terms present in \lsco crystals, they will be discussed in more detail later in the section \cite{LavrovPRL01}.
\lco is an insulator, but small carrier concentrations in the CuO$_{2}$ planes give rise to metallic behavior of the resistivity at high temperatures\cite{AndoPRL01}.
Moreover, the $dc$ resistivity shows also sizeable anisotropy if measured along the $a$ and $b$ orthorhombic axes.
The relative anisotropy is almost 30\% for x~=~0.01 around the metal-insulator transition and goes beyond this value in x~=~0.03 at low temperatures.
One can also notice at high temperatures a decrease of the resistivity anisotropy with doping from x~=~0.01 to x~=~0.03 and that there is a sign change in this anisotropy around 170~K for x~=~0.03 \cite{AndoPRL02}.
The magnetoresistance can be very large (up to 80\%) at low temperatures \cite{AndoPRL03}.

One can conclude form Fig.~\ref{f13} that detwinned samples show non-negligible effects in transport and magnetization data.
The Zeeman energy in finite external magnetic fields becomes comparable with the spin-anisotropy induced gaps and this will influence the low temperature thermodynamics.
As for the intrinsic ground state properties at small dopings, inelastic neutron scattering (INS) argues that there are changes in the low frequency magnetic scattering (45$^{\circ}$ rotation in the $k$ space of low energy incommensurate magnetic peaks) when superconductivity occurs around $x~=~0.05$ in \lsco \cite{WakimotoPRB99} and also that macroscopic phase separation takes place below $x~=~0.02$ \cite{MatsudaPRB02}.

\begin{figure}[t]
\centerline{
\epsfig{figure=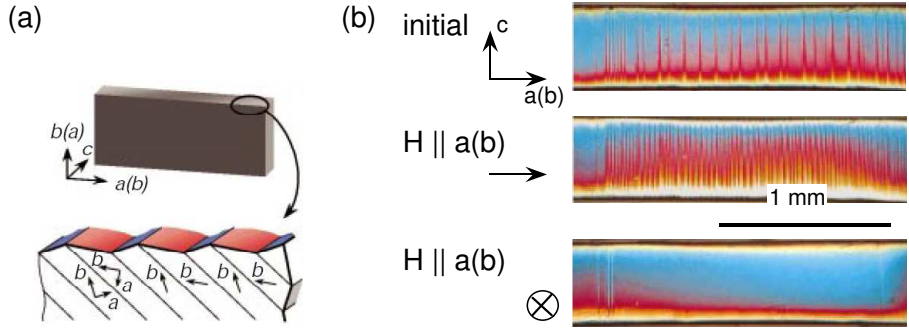,width=120mm}
}
\caption{
(a) A 3D picture with a twinned x~=~0.01 \lsco crystal whose top surface is parallel to the $c$-axis.
The blue and red areas correspond to different twins, $(ac)$ or $(bc)$ domains.
(b) The upper figure is an image of the top crystal surface in zero external magnetic field taken with an optical microscope.
Vertical stripes represent the $(ac)$ and $(bc)$ LTO domains as in panel (a).
The middle and bottom images show the structural changes occurring with an application of an external field of about H~=~14~T.
One can see that the $b$ orthorhombic axis follows the direction of $\vec{H}$. 
}
\label{f14}
\end{figure}

All the above constitute general arguments for a detailed high energy resolution study of long wavelength spin excitations as a function of doping and temperature.
Even more interesting is a recent magnetic field experiment done at room temperature in $x~=~0.01$ \lsco.
The main result of the experiment is shown in Fig.~\ref{f14} and it says that the $b$ orthorhombic axis follows the direction of the applied field \cite{LavrovNature02}.
So magnetic fields of about 10-14~T are able to produce structural changes and detwin the \lsco crystal.
The switch of the crystallographic axes is reversible and can be monitored by using a regular optical microscope.
It is worth noting that 300~K is roughly about 100~K above the 3D long range AF ordering temperature in 1\% doped crystals.
Two interesting points can be mentioned in this regard.
One is that there is strong spin-lattice interaction in this material.
The other one is related to the coupling of the spins to the external field.
While magnetic field induced structural changes are easier to be understood in ferromagnetic crystals because the net magnetic moment can provide a substantial coupling to the external field, the fact that these effects take place in a AF system makes \lsco a unique compound.
The rotation of the orthorhombic axes can be also observed in $dc$ resistivity or magnetic susceptibility measurements by monitoring the changes in the anisotropic properties shown in Fig.~\ref{f13} as a function of the direction of the applied external magnetic field $\vec{H}$.

These results highlight the importance of a magnetic field study of the low energy magnetism in low doped \lsco.
We believe that the field induced spin ordering we observe at temperatures up to 300~K in samples displaying long range AF order is related to the effects shown in Fig.~\ref{f14}.

\subsection{Low Energy Magnetism in Detwinned \lsco with ($0 \leq x \ \leq 0.03$)}

{\bf (A) The 2D Heisenberg antiferromagnets and effects of inter-layer coupling.}
CuO$_{2}$ planes form 2D square lattices and the $S = 1/2$ Cu spins interact antiferromagnetically \emph{via} the intermediate O atoms.
The nearest neighbor super-exchange $J$ takes place along the 180$^{\circ}$ Cu-O-Cu bonds and it has a value of approximately 140~meV \cite{ColdeaPRL01}.
The inter-layer correlations are weak for two reasons: on one hand the spacing between the layers is large and on the other hand the magnetic interaction along this direction is frustrated.
So in the first approximation the spin dynamics (especially in the paramagnetic phase) will be dominated by the properties of a 2D isotropic Heisenberg antiferromagnet.
The starting Hamiltonian to characterize these systems is then:
\begin{equation}
\hat{H}_{2D} = \sum_{<i,j>} J_{ij} \vec{S}_{i} \cdot \vec{S}_{j} 
\label{e11}
\end{equation}
where $\vec{S}_{i}$, $\vec{S}_{j}$ are spins on the sites $i$ and $j$ and $J_{ij} = J \approx 140$~meV when $<i,j>$ corresponds to a pair of nearest neighbor (NN) spins.

The spin-spin correlation function $\xi(T)$ is one of the fundamental parameters characterizing the paramagnetic state.
This quantity is extracted from an equation relating the average staggered magnetization to the inter-site distance, of the type: $< \vec{S}_{i} \cdot \vec{S}_{j} > \ \ \propto \ \ \exp(- r_{ij} / \xi)$.
Here $r_{ij}$ is the distance between the sites $i$ and $j$.
Continuum field theory predicts in the paramagnetic phase a spin-spin correlation length given by: $\xi(T) \propto \frac{c}{2 \pi \rho_{s}} \exp[\frac{2 \pi \rho_{s}}{k_{B} T}]$ \cite{ChakravartyPRB89}.
The parameters $c$ and $\rho_{s}$ are for the spin-wave velocity and spin stiffness respectively.
This correlation length diverges as $T \rightarrow 0$ leading to true long range magnetic order only at zero temperature.
This microscopic result is consistent with a theorem showing rigorously that at any finite temperature a 1D or 2D isotropic Heisenberg model with finite-range interactions can be neither ferromagnetic nor antiferromagnetic \cite{MerminPRL66}.
However, in agreement with theoretical predictions, neutron scattering measurements show that the number of correlated spins within the 2D CuO$_{2}$ planes is substantial even at high temperatures.
For example, at 500~K which is about 200~K above the 3D ordering temperature in \lco, $\xi$ is of the order of 50~\AA\ \cite{KeimerPRB92}, approaching values of 200~-~300 lattice constants around T$_{N}$.

A small interlayer coupling $J_{\perp}$ pushes the N\'{e}el ordering temperature to finite values but does not affect significantly the 2D magnetic correlations.
It is believed that the magnitude of the inter-layer exchange is very small, $J_{\perp} \approx 10^{-5} J < 0.02$~K \cite{KastnerRMP98,ChakravartyPRB89,ThioPRB88}.
In spite of such a small perpendicular exchange, the AF ordering temperatures are quite high and this is due to the large in-plane correlation lengths.
Agreement in terms of the order of magnitude for T$_{N}$ using the above value for $J_{\perp}$ can be obtained simply by comparing the thermal and magnetic energies in:
\begin{equation}
k_{B} T_{N} \approx J_{\perp} (m_{s} S)^{2} \left[ \frac{\xi(T_{N})}{a}\right ]^{2} 
\label{e12}
\end{equation}
where $m_{s}$ is the sublattice magnetization in units of $g \mu_{B}$ and $[\xi(T) / a]^{2}$ is proportional to the number of 'ordered' spins in each CuO$_{2}$ plane.
It should be noted that in the HTT phase of \lsco every Cu atom has eight nearest neighbors in the adjacent planes (four above and four below its own CuO$_{2}$ plane).
Due to symmetry, the super-exchange is almost exactly cancelled and the effective $J_{\perp}$ is even smaller than $10^{-5} J$.
It is the distortion associated with the LTO phase, see Fig.~\ref{f12} which partially lifts this degeneracy giving rise to a reasonably sized, although very small, inter-layer exchange.

Eq.~(\ref{e12}) leaves an open question: how to reconcile similar 3D ordering temperatures (T$_{N}$'s typically in the range between 200 and 300~K) for various layered Cu-O based materials (examples are \lsco, \scocl, \ybco, Nd$_{2}$CuO$_{4}$ or Pr$_{2}$CuO$_{4}$, see Refs.~\cite{KastnerRMP98,MatsudaPRB90}) with rather different exchange paths and accordingly values of $J_{\perp}$ that can be quite far apart.
For instance in \scocl the CuO$_{2}$ planes are exactly flat so it is expected that the cancellation of terms because of the inter-layer frustration would decrease $J_{\perp}$ by another few orders of magnitude, requiring anomalously high $\xi (T_{N})$ in order to satisfy Eq.~(\ref{e12}).
It has been suggested that the 3D ordering temperature T$_{N}$ follows immediately after a 2D Kosterlitz-Thouless (KT) phase transition at T$_{KT}$ due to the in-plane spin anisotropy of the $XY$ type which characterizes all the above mentioned AF materials.
It was found that T$_{KT}$ is appreciable, $\approx 0.25 J / k_{B}$, and quite insensitive to the magnitude of the in-plane anisotropy \cite{MatsudaPRB90}.
This would explain the magnitude as well as the similarity between the measured T$_{N}$'s in various Cu-O based 2D AF's.

\begin{figure}[t]
\centerline{
\epsfig{figure=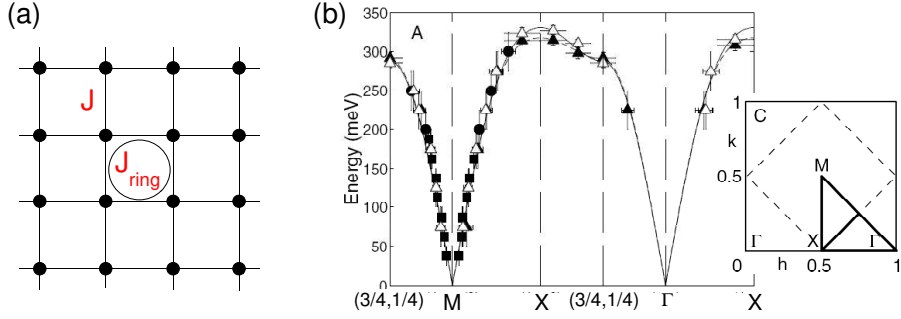,width=120mm}
}
\caption{
(a) The 2D square lattice.
The nearest neighbor superexchange is denoted by $J$ and the circle represents a higher order term considered important for the cuprates, the ring exchange $J_{ring}$.  
(b) The main panel shows the dispersion of the magnetic excitations in \lco at T~=~10~K along different directions in the 2D Brillouin zone (shown in the lower left panel).
The inelastic neutron scattering data is from Ref.~\cite{ColdeaPRL01}.
Squares, circles and triangles correspond to incoming neutron energies of 250, 600 and 650~meV respectively.
The solid line is a fit to the data as described in the text. 
}
\label{f15}
\end{figure}
What does the excitation spectrum of a 2D AF ordered square lattice look like?
Within the spin-wave approximation the excitations are coherent transverse oscillations of the ordered moments.
Taking into account only the nearest neighbor exchange $J$ the wavevector dependent spin-wave energies are given by
\begin{equation}
\omega(k) = z S J \sqrt{1 - \gamma^{2}_{k}} \ \ \ \mathrm{with} \ \ \ \gamma_{k} = \frac{\cos(k_{x}) + \cos(k_{y})}{2}
\label{e13}
\end{equation}
where $S = 1/2$ is the total spin and $z = 4$ is the number of nearest neighbors for the simple square lattice \cite{SandvikPRB98}.
Note that in the 2D isotropic Heisenberg AF lattice there will be two degenerate acoustic spin-wave branches.
In Fig.~\ref{f15} we show relatively recent INS results for the spin-wave dispersion up to high energies.

While the dispersion predicted by the nearest neighbor isotropic Hamiltonian reproduces qualitatively the experimental results, there are discrepancies at high energies.
One can note that along the AF zone boundary we have $k_{x} + k_{y} = \pi$ and this implies that the spin-wave energy along this line is a constant given by $2 J$.
The experimental data in Fig.~\ref{f15} shows that there is substantial dispersion for instance along the $(\pi,0)$ to $(3 \pi / 2, \pi / 2)$ line.
The authors resolve this discrepancy by including higher order spin interactions.
In particular, the most prominent term is due to $J_{ring}$ which corresponds to a spin exchange around a square plaquette as shown in Fig.~\ref{f15}a.
Quantitatively, from the fit to the experimental data which includes quantum corrections \cite{SinghPRB89} (the solid line in Fig.~\ref{f15}b), this term turns out to be as high as 41\% of the nearest neighbor $J$ at low temperatures, $J_{ring} \approx 61$~meV, and it is about twenty times larger than the second and third nearest neighbor exchanges \cite{ColdeaPRL01}!
In support for such a claim we note that a large value of $J_{ring}$ was needed to explain the dispersion of the elementary triplet excitations in two-leg ladder materials. 
The same $J_{ring}$ seems also to improve the results concerning the large absorption frequency range in which the phonon induced two-magnon excitation is thought to be observed in 2D insulating cuprates \cite{LorenzanaPRL99}.

{\bf (B) In-plane magnetic anisotropies.}
There are two dominant in-plane magnetic anisotropies characterizing each CuO$_{2}$ plane.
In general these terms, arising as a result of spin-orbit coupling, connect the spin space to the real space and can be sometimes described in terms of effective magnetic interactions.
One of these interactions is the $XY$ exchange anisotropy term mentioned above in connection to the 3D N\'{e}el ordering and its origin is in the layered structure of the cuprates, i.e. it has nothing to do with the buckling of the CuO$_{2}$ plane in the LTO phase.
Due to to the $XY$ term the NN spin-exchange interaction in Eq.~(\ref{e11}) changes to $(J + \alpha) (S^{x}_{i} S^{x}_{j} +  S^{y}_{i} S^{y}_{j}) + J S^{z}_{i} S^{z}_{j}$.
Because $\alpha > 0$ the classical configuration giving the minimum energy is one with all the spins lying in the $(ab)$ plane. 
The other important anisotropy term is due to the antisymmetric Dzyaloshinskii-Moriya (DM) interaction and  it has the form: $\vec{d}_{ij} \cdot (\vec{S}_{i} \times \vec{S}_{j})$ where $\vec{d}_{ij}$ is the DM vector \cite{DzyaloshinskiiJPCS58,MoriyaPR60}.
The two-spin classical ground state configuration for this interaction considered alone is one with $\vec{S}_{i} \perp \vec{S}_{j} \perp \vec{d}_{ij}$.

The balance of these terms determines the equilibrium position of the spins.
These anisotropy terms are expected to be much smaller than $J$ and can be quantitatively determined from the energy of the spin-waves in the long wavelength limit as will be discussed in the next section.
A few words are in order about the DM term.
Due to the existence in the LTO phase of a $C_{2}$ (rotation by 180$^{\circ}$) symmetry axis which passes through in-plane O atoms and is perpendicular to the $(ab)$ surface, the DM vector between two adjacent Cu atoms has to satisfy $\vec{d} \perp \hat{c}$ \cite{MoriyaBook}.
\begin{figure}[b]
\centerline{
\epsfig{figure=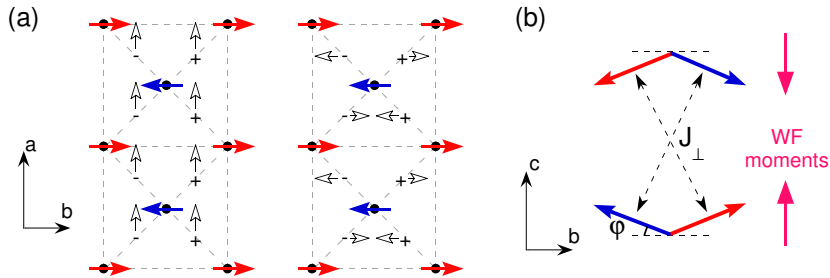,width=110mm}
}
\caption{
(a) Top views of the CuO$_{2}$ planes corresponding to the two possible configurations of the DM interactions (black arrows).
Cu atoms are represented by black dots; their spins (lying mainly in the $(ab)$ plane because of the $XY$ spin anisotropy and forming a two-sublattice AF structure) are shown by red and blue arrows.
By '+' and '-' signs we denote the intermediate O atoms which are below and above the plane of the paper, see Fig.~\ref{f12}.
(b) Cartoon with the 3D four-sublattice configuration of spins seen in the $(bc)$ plane and the weak ferromagnetic moments (proportional to the canting angle $\phi$) of each CuO$_{2}$ plane due to the Dzyaloshinskii-Moriya interaction $\vec{d} \parallel \hat{a}$ shown in the left of panel (a). 
}
\label{f16}
\end{figure}
The symmetry elements of the $Bmab$ space group associated to the LTO phase allow the DM vectors $\vec{d}_{ij}$ to form the configurations shown in Fig.~\ref{f16}a.
Once a convention is made that the order of spins in the vector product of the $\vec{d}_{ij} \cdot (\vec{S}_{i} \times \vec{S}_{j})$ term is always from a given sublattice to the other, it can be noted that there are two possible arrangements for the DM vectors: one involving $\vec{d} \parallel \hat{a}$ and the other one in which the DM vectors are parallel to the $b$-axis but have alternating signs.  

The effective two-dimensional spin Hamiltonian and the associated free energy density at T~=~0~K which takes the $XY$, DM terms as well as an external field into account can be written as:
\begin{eqnarray}
\hat{H} = \sum_{<i,j>} [ (J + \alpha) (S^{x}_{i} S^{x}_{j} +  S^{y}_{i} S^{y}_{j}) + J S^{z}_{i} S^{z}_{j} + \vec{d}_{ij} \cdot ({\bf S}_{i} \times {\bf S}_{j})] - \vec{H} \sum_{i} \vec{S}_{i}
\label{e14}
\\
f = z (J + \alpha) (M^{x}_{1} M^{x}_{2} +  M^{y}_{1} M^{y}_{2}) + z J M^{z}_{1} M^{z}_{2} + z \vec{d} \cdot (\vec{M}_{1} \times \vec{M}_{2}) - \vec{H} (\vec{M}_{1} + \vec{M}_{2})
\label{e15}
\end{eqnarray}
In Eq.~(\ref{e14}) the sum runs over the nearest neighbors.
In Eq.~(\ref{e15}) $z$ is the number of nearest neighbors and $M_{1,2}$ are the sublattice magnetizations.
The 'thermodynamic' Dzyaloshinskii vector $\vec{d}$ in Eq.~(\ref{e15}) is given in terms of the 'microscopic' Moriya terms $\vec{d}_{ij}$ in Eq.~(\ref{e14}) by $\vec{d} = (1 / z) \sum_{NN} \vec{d}_{ij}$ and from Fig.~\ref{f16} one can infer that $\vec{d} \parallel \hat{a}$ \cite{ShekhtmanPRL92}.
The relative strength of the DM terms corresponding to the two configurations in Fig.~\ref{f16}a is determined by microscopic parameters.
It is interesting to note here a point made by the authors of Ref.~\cite{ShekhtmanPRL92}, i.e. that the identification of $\vec{d}$ to the microscopic $\vec{d}_{ij}$ is a non-trivial problem in the sense that a difference between them is a necessary condition for the existence of an observable weak ferromagnetism (WF) with a specific value of the net WF moment.
In other words, although $\vec{d}$ is parallel to the $a$-axis, it is required that the DM vectors in both configurations shown in Fig.~\ref{f16}a are finite.
If on one hand only vectors $\vec{d}_{ij} \parallel \hat{b}$ are considered (frustrating interaction), then $\vec{d} \equiv 0$ and the spins order antiferromagnetically without any WF moment.
On the other hand if one takes into account only vectors $\vec{d}_{ij} \parallel \hat{a}$ (non-frustrating interaction) the classical ground state cannot be characterized as ferromagnetic because it consists of a manifold of degenerate configurations having a net WF moment ranging continuously from zero to some finite value \cite{ShekhtmanPRL92}.

The equilibrium position of the spins in zero external field is shown in Fig.~\ref{f16}b.
For a 2D plane this can be obtained from the minimization of the free energy in Eq.~(\ref{e15}) with respect to the angles between the magnetizations and crystallographic axes with the constraint $m = |\vec{M}_{1}| = |\vec{M}_{2}|$.
The canting angle is given by $\tan(2 \varphi) = 2 d / (2 J + \alpha)$ and since $d \ll J$ (in reality $\varphi < 0.5^{\circ}$) the net WF moment of each plane is approximately $M_{F} \approx 2 m \varphi = 2 d m / (2 J + \alpha)$.
Here $m$ is the sublattice magnetization.
The interaction $J_{\perp}$ does not significantly change this angle.

{\bf Long wavelength spin-wave excitations}
On general grounds, from Eq.~(\ref{e15}) one can say the following about the behavior the spin-wave modes in the long wavelength limit: (1) if $\alpha = d = 0$ there will be two acoustic modes; (2) if $\alpha \neq 0$ and $d = 0$ or $\alpha = 0$ and $d \neq 0$ there will be one acoustic and one gapped spin-wave branch; (3) if $\alpha \neq 0$ and $d \neq 0$ both spin-wave branches will be gapped.
This is because unless we are in case (3), there is a global continuous symmetry which is broken at the AF transition due to the ordering of the magnetic moments (the gapless branches are typical Goldstone modes).

\begin{figure}[t]
\centerline{
\epsfig{figure=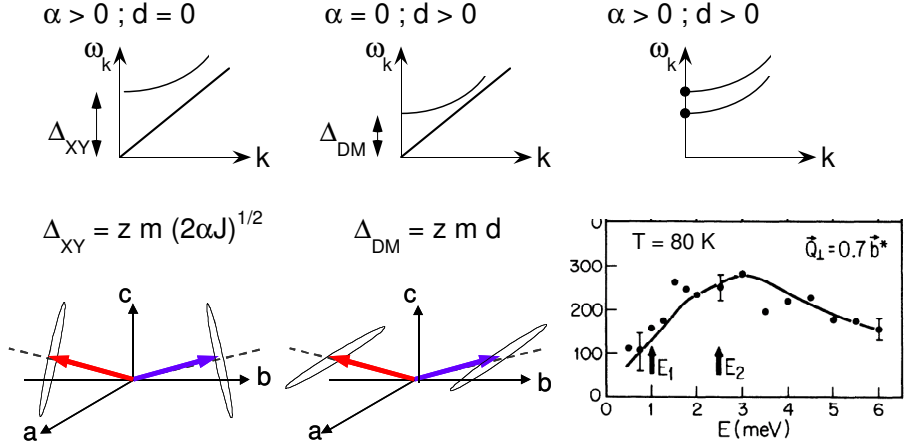,width=120mm}
}
\caption{
The upper row shows schematically the dispersions of the spin-wave branches in the limit $k \rightarrow 0$ if: only the $XY$ (left), only the DM (middle) and both of these anisotropy terms are present (right).
The lower row shows the pattern of oscillating magnetizations corresponding to the $XY$ (left) and $DM$ (middle) spin gaps at $k = 0$ as well as the INS experimental data from Ref.~\cite{PetersPRB88} showing these two gaps in \lco (E$_{1}$ and E$_{2}$ correspond to the DM and $XY$ gaps respectively).
}
\label{f17}
\end{figure}
The situation described above is shown schematically in Fig.~\ref{f17}.
One can intuitively understand how the spin gaps look like at a classical level by solving the equations of motion:
\begin{equation}
\beta \frac{\partial \vec{M}_{j}}{\partial t} \ = \ \vec{M}_{j} \times \nabla_{\vec{M}_{j}} f \ \ \ \ 
j = 1,2
\label{e16}
\end{equation}
where $\beta$ is a constant related to the Bohr magneton and $f$ is the free energy from Eq.~(\ref{e15}).
The equilibrium condition $\nabla_{\vec{M}_{j}} f = 0$ gives the ground state shown in Fig.~\ref{f16}b.
Linearizing the equations of motion from (\ref{e16}) around equilibrium and choosing oscillatory solutions for the obtained set of homogeneous equations one can get (to first order in anisotropy terms) the following energies corresponding to the $XY$ and DM gap respectively:
\begin{equation}
\omega_{XY} = z \ m \ \sqrt{2 \alpha J} \ \ \  \mathrm{and} \ \ \ \omega_{DM} = z \ m \ d
\label{e17}
\end{equation}
With $z = 4$, $m = 1/2$ and taking $J = 145$~meV \cite{ColdeaPRL01} one can calculate from Eq.~(\ref{e17}) the anisotropy parameters $\alpha$ and $d$ if $\omega_{XY}$ and $\omega_{DM}$ are known.
If the quantum corrections for the spin-wave velocity are taken into account \cite{SinghPRB89} the expressions for the gap energies become $\omega_{XY} = 2.34 \sqrt{2 \alpha J}$ and $\omega_{DM} = 2.34 d J$ \cite{KastnerRMP98}.
The ellipses shown in Fig.~\ref{f17} are very elongated, the ratio of their small and big axes being essentially given by ratios of the anisotropy parameters with respect to the large super-exchange $J$.
This is why in the literature the $XY$ mode (which corresponds to the precession of the net WF moment around the $c$-axis) is also called the out-of-plane gap while the DM mode (which corresponds to the $c$-axis oscillations of the WF moment) is called the in-plane gap.

In Fig.~\ref{f17} we also show the low energy INS measurements in \lco of Peters \emph{et al.} \cite{PetersPRB88}.
The dots are the experimental data and the solid line is a fit using the spin-wave approximation convoluted with the experimental resolution.
The energies of these two gaps are shown by arrows.
The most direct way to check the magnetic nature of these modes is to apply an external magnetic field which has not been done so far.
It is also desirable that such a study be performed with a higher energy resolution probe.

\subsection{Magnetic Field, Temperature and Doping Dependence of the Dzyaloshinskii-Moriya Gap in \lsco ($0 \leq x \ \leq 0.03$)}

\begin{figure}[b]
\centerline{
\epsfig{figure=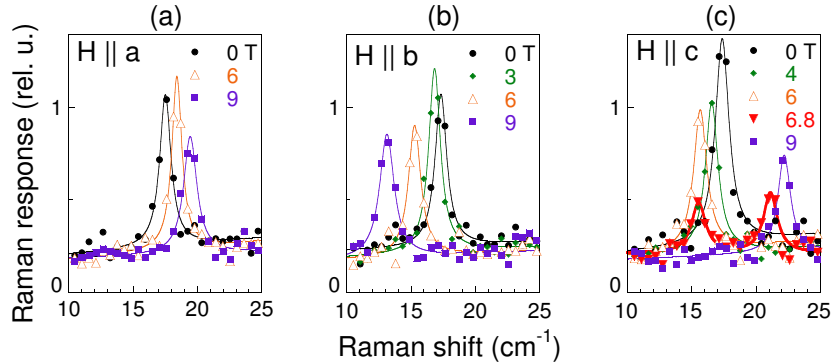,width=110mm}
}
\caption{
T~=~10~K magnetic field dependence of the Dzyaloshinskii-Moriya gap in \lco.
The Raman spectra are taken in $(RL)$ polarization.
The dots are experimental data and the solid lines are Lorentzian fits. 
}
\label{f18}
\end{figure}

\begin{figure}[t]
\centerline{
\epsfig{figure=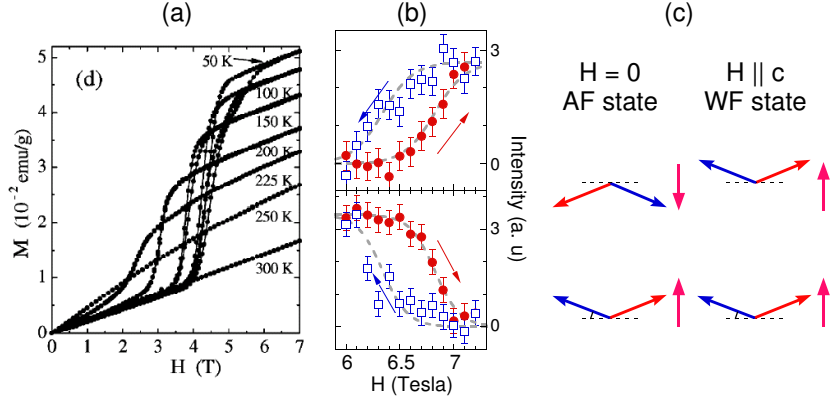,width=110mm}
}
\caption{
(a) Magnetization as function of magnetic field parallel to the $c$-axis for several temperatures.
Data are from Ref.~\cite{AndoPRL03} and they are taken from a x~=~0.01 \lsco crystal.
(b) Hysteretic loops of the 21~\cm-1 (upper panel) and 15~\cm-1 (lower panel) modes from Fig.~\ref{f18}c at the WF transition.
(c) The spin arrangements in the 3D AF and WF states.
}
\label{f19}
\end{figure}

{\bf (A) Field dependent Dzyaloshinskii-Moriya gap in \lco.}
Fig.~\ref{f18} shows 10~K Raman spectra taken from 93\% detwinned \lco crystal using circular $(RL)$ polarization. 
A sharp resonance is seen in zero field at 17~\cm-1.
This excitation disperses continuously upwards (downwards) for $\vec{H} \parallel \hat{a}$ ($\vec{H} \parallel \hat{b}$) axes.
For $\vec{H} \parallel \hat{c}$, Fig.~\ref{f18}c, the mode disperses downwards until the magnetic field reaches the value $H_{WF} \approx 6$~T.
At this point a transition to the WF state takes place.
Initially observed in magnetic field dependent magnetization data \cite{ThioPRB88} this first order transition could be also studied by neutron scattering measurements \cite{KastnerPRB88} because due to the change in magnetic symmetry the scattering form factors allowed new Bragg peaks.
In the 6~-~7~T range the resonance remains around 15~\cm-1 but decreases in intensity and we observe a concomitant appearance of another feature around 21~\cm-1.
The 6.8~T spectrum in Fig.~\ref{f18}c shows clearly the coexistence of AF and WF states.
Recent magnetization data \cite{AndoPRL03} and with a cartoon comparing the spin configuration in zero field and in the WF state are shown in Fig.~\ref{f19}.
The hysteretic loops in the magnetic field dependent magnetization correspond to the hysteretic loops of the 15 and 21~\cm-1 modes shown in Fig.~\ref{f18}b.
This is in turn very similar to the behavior of the (100) and (201) magnetic Bragg peaks~\cite{KastnerPRB88}, reflecting the dynamics of magnetic domains in the presence of small crystalline imperfections. 

\begin{figure}[t]
\centerline{
\epsfig{figure=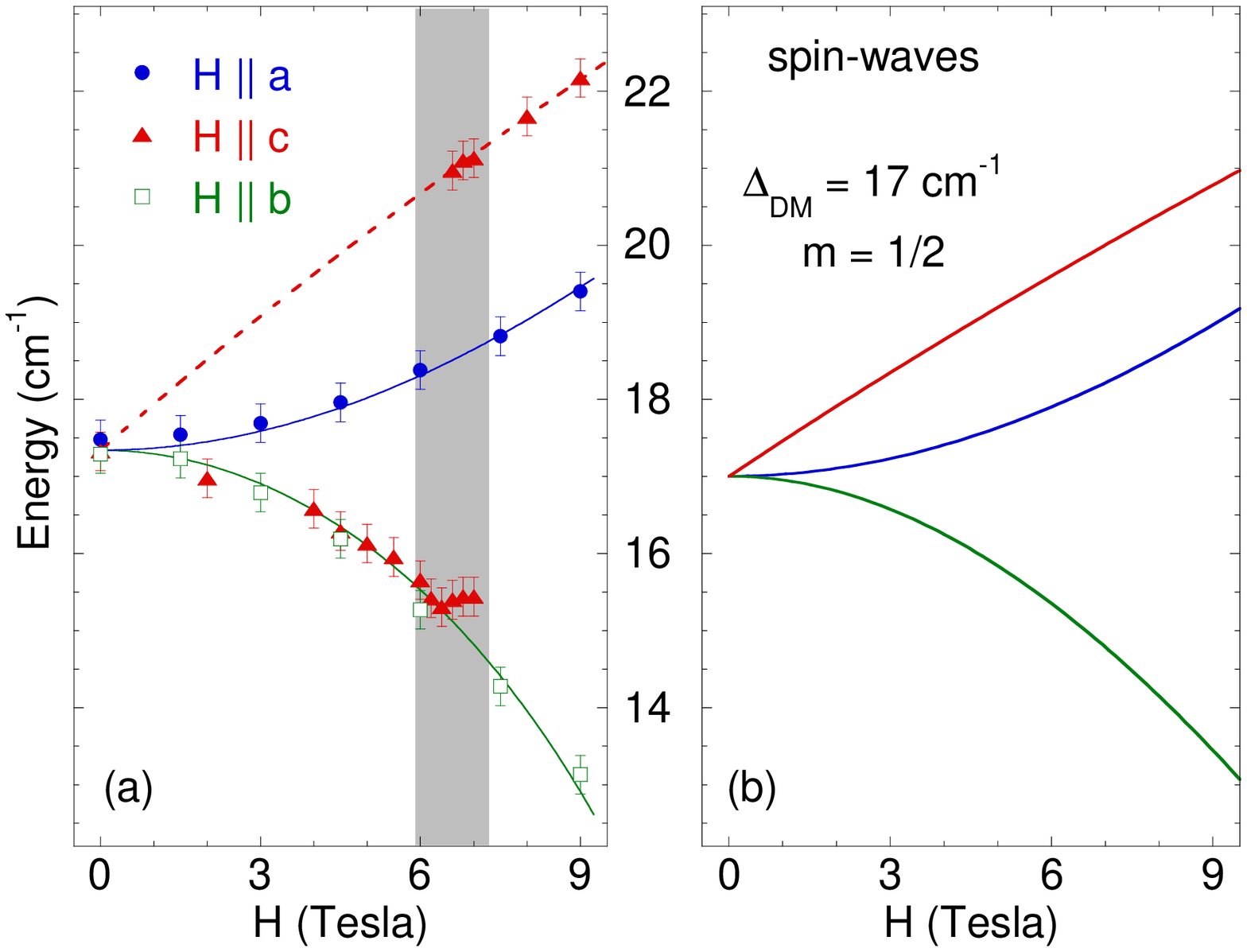,width=90mm}
}
\caption{
(a) Magnetic field dependence of the DM gap in \lco for $\vec{H} \parallel \hat{a}$ (filled blue circles), $\vec{H} \parallel \hat{b}$ (empty green squares) and $\vec{H} \parallel \hat{c}$ (filled red triangles).
The $\vec{H} \parallel \hat{c}$ data show the transition to the WF state depicted in Fig.~\ref{f19}c.
The shaded area corresponds to the region of coexistence of the AF and WF states.
The continuos lines are fits as described in the text.
(b) Results of a 2D spin-wave calculation for the DM gap dispersion using Eqs.~(\ref{e15}) and (\ref{e16}) and assuming a fully ordered moment on Cu sites. 
}
\label{f110}
\end{figure}

The energies of the 17~\cm-1 resonance as a function of magnetic field are plotted in Fig.~\ref{f110}a.
The expressions for the fitting functions and the values for the fitting parameters used for the data in this figure are summarized in the following.
If $\vec{H} \parallel \hat{a}$ and $\vec{H} \parallel \hat{b}$ $\ \ \ \Rightarrow \ \ \ \Delta_{H} = \sqrt{\Delta_{DM}^{2} + \gamma H^{2}}$ with $\Delta_{DM} = 17.35 \pm 0.25$~cm$^{-1}$ and $\gamma_{H \parallel a} = 0.96$ and $\gamma_{H \parallel b} = -1.65$~(cm T)$^{-2}$.
If $\vec{H} \parallel \hat{c}$ and $H \geq H_{WF}$ $\ \ \ \Rightarrow \ \ \ \Delta_{H} = \sqrt{\Delta_{DM}^{2} + \beta H}$ with $\beta = 22.6$~cm$^{-2}$T$^{-1}$.
The quadratic dependence in the first two cases can be understood because the spin re-arrangement in finite magnetic fields is independent of the directions parallel to the $a$ or $b$ axes along which the field is applied.
This is not the case if $\vec{H} \parallel \hat{c}$ and the system is in the WF state, see Fig.~\ref{f19}c.
Note that if $\vec{H} \parallel \hat{c}$ but $H \leq H_{WF}$ one observes again a quadratic dispersion with field.
Moreover, the similar field dispersion for $\vec{H} \parallel \hat{b}$ versus $\vec{H} \parallel \hat{c}$ ($H < H_{WF}$) seen in Fig.~\ref{f110}a is intriguing because this degeneracy does not follow from the model of Eq.~(\ref{e15}) but it rather suggests rotational symmetry with respect to the $a$-axis.

Confirmation that the 17~\cm-1 (in zero field) resonance observed in Fig.~\ref{f18} is the DM spin-wave gap comes from a 2D semiclassical spin-wave calculation.
Assuming a fully ordered moment on Cu sites ($m = 1/2$), a zero field DM gap $\Delta_{DM} = 17$~\cm-1 and minimizing Eq.~(\ref{e15}), one can obtain the dispersions of the $k = 0$ DM gap for the three directions of the applied field.
The results are shown in Fig.~\ref{f110}b and the reasonable agreement with the experimental data allows one to assign this excitation to the DM interaction induced spin gap.
Two comments on Fig.~\ref{f110}b.
The first is that the 2D calculation can account only for the situation where the two sublattices in adjacent CuO$_{2}$ planes 'respond similarly' to the external field.
This is the case for $\vec{H} \parallel \hat{a}$, $\vec{H} \parallel \hat{b}$ and $\vec{H} \parallel \hat{c}$ with $H \geq H_{WF}$ and one can see that in all these cases the theoretical predictions agree with the experiment.
If $\vec{H} \parallel \hat{c}$ and $H \leq H_{WF}$ the 2D approximation clearly breaks down and Eq.~(\ref{e15}) cannot be used in this region.
The second comment is just a remark that it is surprising that a semi-classical spin-wave calculation as shown in Fig.~\ref{f110} is able to reproduce with relatively good accuracy the experimental data in a low spin system.
This is in view of the expectation that such an approximation is valid to order $1 / S$ \cite{AndersonPR52} which is not a 'small' number for $S = 1/2$.
One may conclude from here that in order to explain the low energy spin dynamics in undoped 2D cuprates one does not need to go beyond a semiclassical approximation.

We believe that the magnetic field dependent data shown in Figs.~\ref{f18} and \ref{f110} may also be relevant for a quantitative estimation of higher order spin interactions which are thought to be important in cuprates.
One example is the ring exchange, see Fig.\ref{f15}, and it would be interesting to check the influence of $J_{ring}$ on the DM gap energy and possible renormalization effects on its magnetic field dependence.
An example of a system where substantial effects of $J_{ring}$ on the spin-gap were pointed out is that of two-leg spin ladders. 
Using the expression $\Delta_{DM} = 2.34 d$ we can extract for \lco the value $d = 0.92 \pm 0.013$~meV. 

\begin{figure}[t]
\centerline{
\epsfig{figure=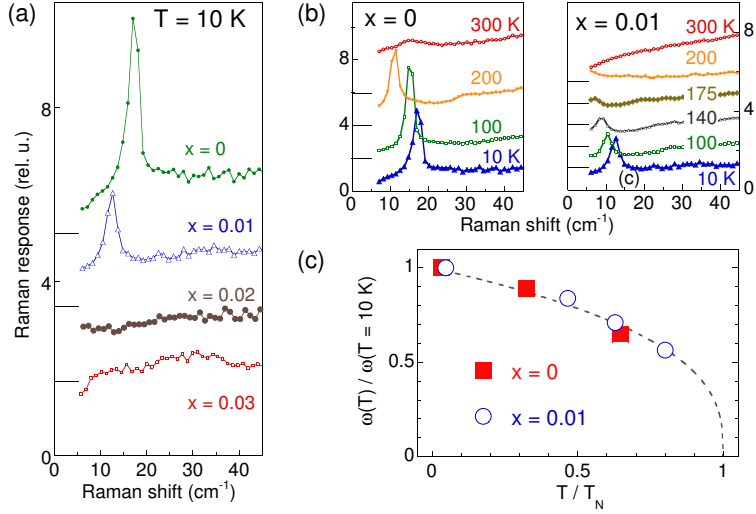,width=100mm}
}
\caption{
Doping and temperature dependence of the DM gap in \lsco in $(ab)$ polarization for zero applied field.
(a) T~=~10~K data for x~=~0, 0.01, 0.02 and 0.03.
(b) Temperature dependent spectra for x~=~0 and 0.01.
(c) The energy of the DM gap with respect to the value at 10~K, $\omega (T) / \omega (10 K)$, as a function of $T / T_{N}$.
The dashed line is a guide for the eye.
}
\label{f111}
\end{figure}

{\bf (B) Doping and temperature effects on the Dzyaloshinskii-Moriya gap.}
The results of doping and temperature on the DM gap are summarized in Fig.~\ref{f111}.
The N\'{e}el temperatures for the x~=~0 and 0.01 crystals studied here are 310 and 215~K respectively.
The 1 and 3\% Sr doped crystals were detwinned in proportion of 98 and 97\%.
In Fig.~\ref{f111}a we show the gap as a function of doping at 10~K.
The 17~\cm-1 resonance in the undoped crystal seen in the B$_{1g}$ orthorhombic channel becomes weaker in intensity, remains as sharp as in \lco and softens to 12.5~\cm-1 for x~=~0.01, an energy 30\% smaller compared to what we see for x~=~0.
We note also the absence of the DM mode in x~=~0.02 and 0.03 \lsco crystals.

The doping dependence shows that the the DM mode is present at low temperatures only in the long range AF ordered region of the phase diagram (Fig.~\ref{f11}).
This behavior is somehow surprising because one would expect to see for 2 or 3\% doping at least a broadened feature in view of the large 2D magnetic correlations just outside the AF ordered phase.
Such fluctuations are not observed in our data and this suggests that the presence of the long wavelength DM excitation is related to the existence of a true 3D order.
A point discussed in the preceding section was related to the fact that it is the orthorhombicity which generates the DM interaction.
The decrease of almost 30\% in its energy from x~=~0 to 0.01 is much more pronounced compared with the decrease in orthorhombicity and this relates this renormalization effects to a strong sensitivity on hole doping and points to a considerable renormalization of the ordered Cu moment at only 1\% hole doping.
Our data suggest that the antisymmetric interaction is strongly competing with other sources of disorder in the magnetic system and we suggest that this is most likely due to the frustration effects and the associated spin distortions induced by hole doping \cite{AharonyPRL90GoodingPRB97}.
A last point we make regarding Fig.~\ref{f111}a is in regard to the macroscopic phase separation scenario proposed by the authors of Ref.~\cite{MatsudaPRB02} to take place for $x \leq 0.02$.
If this were true than one would observe in the $x~=~0.01$ crystal two features: one corresponding to the undoped region which would be found at 17~\cm-1 and another one corresponding to the region with carrier concentration $c_{h} \approx 0.02$.
The 30\% decrease in energy observed in the $x~=~0.01$ crystal with respect to the undoped case seems to rule out the scenario proposed in Ref.~\cite{MatsudaPRB02}.

As a function of temperature what we see in Fig.~\ref{f111}b-c is that the DM gap softens with raising the temperature and disappears below 5~\cm-1 as we approach the N\'{e}el temperature from below in both $x~=~0$ and 0.01 \lsco samples.
The temperature dependence of the peak energies in the two crystals is shown in Fig.~\ref{f111}c to be similar and points towards a conventional soft mode behavior of this excitation, i.e. both its energy and its intensity approach zero in the limit $T \rightarrow T_{N}$, $T < T_{N}$.
These spectra support the statement made in the previous paragraph that the DM induced gap exists in the narrow (T,$x$) region of the phase diagram from Fig.~\ref{f11} where the AF order is long ranged.
It is possible that the reason it disappears at higher dopings is because the low energy magnetic fluctuations move away from the Brillouin zone center \cite{TranquadaNature04}. 
Interestingly, a broad peak at 300~K is seen around 15~\cm-1 for $x~=~0$.
This peak becomes a kink at 200~K and, as opposed to the conventional behavior of the DM gap, it disappears with further cooling.
It is the purpose of the following section to investigate this excitation.

\subsection{Magnetic Field Induced Spin Ordering in $x = 0$ and $0.01$ \lsco}

\begin{figure}[t]
\centerline{
\epsfig{figure=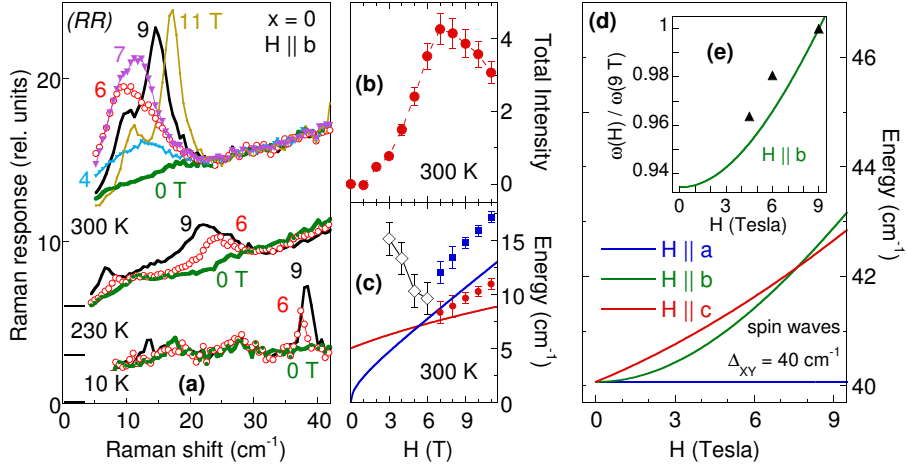,width=120mm}
}
\caption{
(a) Magnetic field dependence of $(RR)$ polarized spectra in \lco at T~=~10, 230 and 300~K.
Data are vertically offset.
(b) Integrated intensities obtained by subtracting the room temperature 0~T spectrum from the finite field data at the same temperature.
(c) The energies of the observed magnetic modes at T~=~300~K.
Empty diamonds show the energy of the broad peak seen for magnetic fields $H \leq 6$~T.
The solid lines in this panel are calculated using eqs.~(\ref{e15}) and (\ref{e16}) assuming $\alpha = 0$, see the text for a discussion.
(d) This panel shows the results of a T~=~0~K spin-wave calculation similar as the one shown in Fig.~\ref{f110} but in this case for an $XY$ gap assumed to be at $\Delta_{XY} = 40$~\cm-1.
(e) Using the calculation from (d), the solid line in the inset shows the magnetic field dependence of the ratio $\omega(H) / \omega(9 T)$ for fields $\vec{H} \parallel \hat{b}$ axis.
The dots represent the same quantity extracted from the T~=~10~K experimental data in panel (a).
}
\label{f112}
\end{figure}

Magnetic field dependent $(RR)$ polarized Raman spectra in \lco at several temperatures in $\vec{H} \parallel \hat{b}$ configuration are shown in Fig.~\ref{f112}a.
At 10~K and in zero external field the Raman spectrum is featureless.
For $H = 6$~T we see a sharp field induced mode (FIM) situated at 37.5~\cm-1 which moves to slightly higher frequency (38.3~\cm-1) for $H = 9$~T.
The triangles in Fig.~\ref{f112}e show the relative energy of this excitation with respect to the value at 9~T.
The extra data point corresponding to the 4.5~T Raman spectrum (not shown for clarity in panel (a)) marks the magnetic field at which the FIM starts to be seen.
We remark only a small hardening (of about 4\%) with magnetic field from 4.5 to 9~T.
At 230~K the FIMs in 6 and 9~T fields are broader than at 10~K.
However, with increasing field the FIM softens gaining spectral weight from the lower energy side.
At 300~K, as long as the field is less than about 6~T, we observe qualitatively similar behavior as at 230~K.
For magnetic fields beyond that value we see the emergence of two independent peaks and both of them harden with further increasing the field.
Fig.~\ref{f112}b we plot the total integrated intensity of the magnetic modes (for T~=~300~K) at a given field, the data showing a maximum around $H = 7$~T, and in panel (c) the symbols denote the position of the FIMs as the magnetic field is swept from 0 to 11~T.

If $\vec{H} \parallel \hat{a}$ or $\vec{H} \parallel \hat{c}$ we do not observe any changes in the $(RR)$ polarized Raman spectra.
Note that the spectra showing the DM gap in Fig.~\ref{f18} were taken in $(RL)$ polarization.
Circular polarizations probe 'good' symmetries if the crystal has a symmetry higher than tetragonal.
Because the orthorhombicity in our samples is small it allows to separate the excitations appearing in these two geometries, but because it is finite we observe small 'leakage' effects.
Their magnitude can be estimated for instance by looking at the small feature corresponding to the DM gap which is found around 6-7~\cm-1 in the H~=~9~T and T~=~230~K spectrum from Fig.~\ref{f112}a.
\begin{figure}[t]
\centerline{
\epsfig{figure=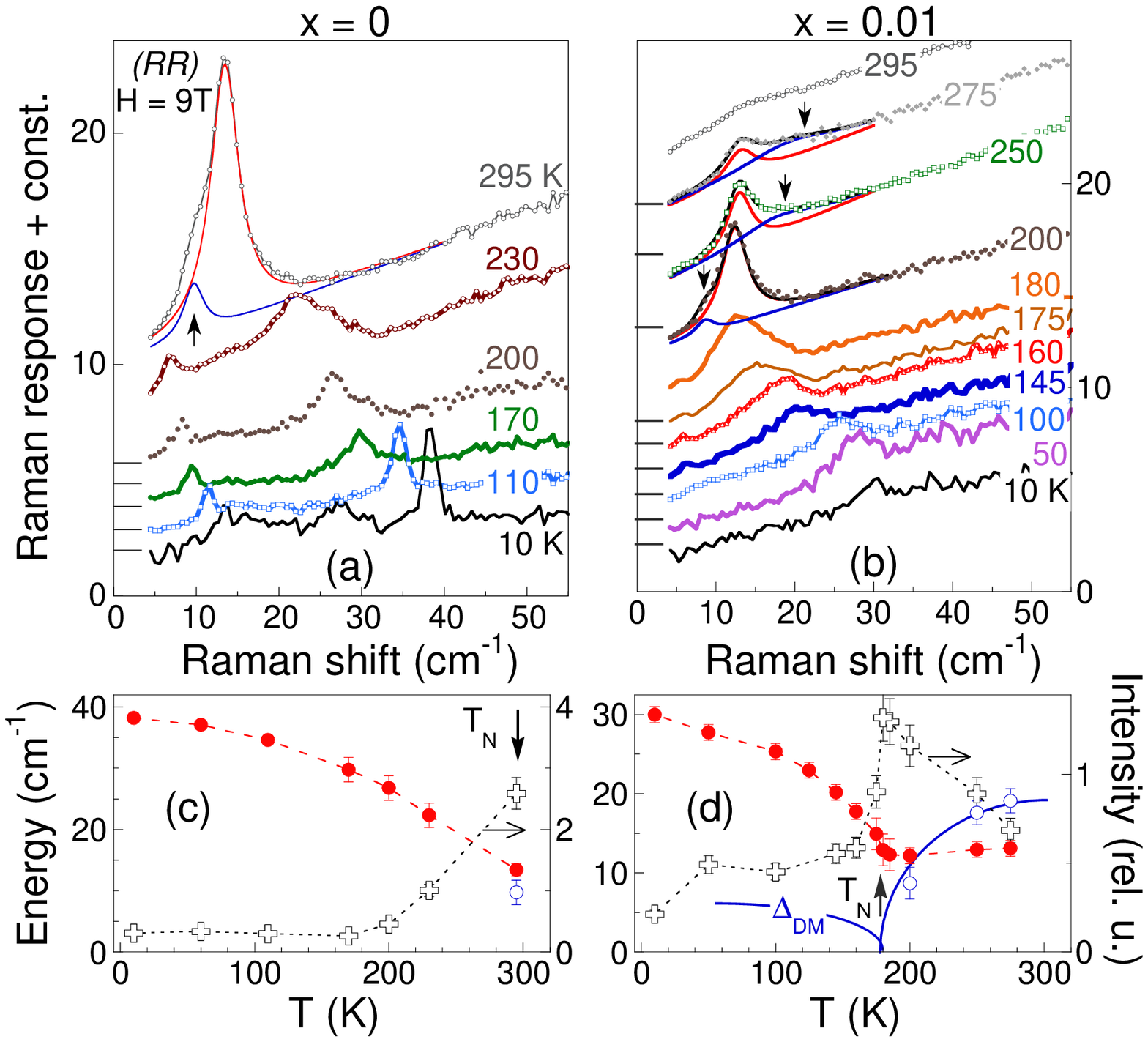,width=110mm}
}
\caption{
Temperature dependence of the field induced mode (FIM) in La$_{2-x}$Sr$_{x}$CuO$_{4}$ for $x = 0$ (left) and $0.01$ (right).
(a-b) Data (vertically offset) in $(RR)$ polarization for $\vec{H} \parallel \hat{b}$ at 9~T.
The continuous lines for T~=~295~K in (a), T~=~200, 250 and 275~K in (b) are two-Lorentzian fits to the data.
(c-d) Variation with temperature of the FIM energies (filled red circles, left scales) and intensities (crosses, right scales).
The empty blue circles correspond to the arrows in (a-b).
The blue lines in (d) are guides for the eye.
We also show by arrows the N\'{e}el temperatures for $\vec{H} \parallel \hat{b}$ at 9~T in the two samples. 
}
\label{f113}
\end{figure}

In \lco the FIMs dynamics marks two events.
The first seems to be a phase transition at 300~K and fields around 6~T.
This is indeed the case because we know that the N\'{e}el temperature in \lco is around 310~K and
that the magnetic susceptibility $\chi_{b}$ shows T$_{N}$ decreasing at a rate of about 1~K/T if the magnetic field is applied parallel to the $b$-axis, as is the case in Fig.~\ref{f112}.
Moreover, the narrow widths of the magnetic excitations above 6~T (2~\cm-1~$\approx$~0.25~meV) at temperatures more than two orders of magnitude higher (300~K~$\approx$~25~meV) argue strongly
for the collective nature of these excitations which correspond to another magnetically ordered state with a well defined gap in the excitation spectrum.
Such a transition is expected from the low temperature data shown in Figs.~\ref{f18} and \ref{f110}, more precisely from the behavior of the DM gap for $\vec{H} \parallel \hat{b}$.
In this configuration we can fit the behavior of the DM gap by $\sqrt{\Delta_{DM}^{2} + \gamma_{b} H^{2}}$ with $\gamma_{b} < 0$.
Extrapolating to higher fields would lead to a collapse of this gap marking a field induced transition.
The second event, a crossover taking place between 230 and 10~K, is reflected in the opposite dispersion with field and different peak widths at these two temperatures.

As for the doping dependence, except for a much weaker intensity (see Fig.~\ref{f113}), we observed the same qualitative behavior in x~=~0.01 \lsco.
The FIM is not seen (in fields up to 9~T) at any temperature for $x \geq 0.02$.
Accordingly, it seems that, like the DM gap, this feature is a characteristic of the phase diagram where long range AF order exists.
As for the DM gap, the reason for its absence at higher dopings could be because the low energy magnetic excitations move away from $k = 0$ concomitant to the development of incommensurate magnetic excitations.
In the following we try to identify the nature of the FIM and field induced transition by looking at the effects of the temperature on the Raman data in magnetic fields.

Fig.~\ref{f113} shows temperature dependent $(RR)$ polarized spectra in a 9~T field $\vec{H} \parallel \hat{b}$ for $x~=~0$ and 0.01.
The data in panel (a) show that the crossover mentioned above (regarding the change in the FIM width and energy dispersion with field) takes place around 150~K.
This is the temperature below which the FIM width narrows.
Fig.~\ref{f113}c shows that the intensity of this excitation increases as we approach T$_{N}$ from below and that around 300~K we observe the splitting due to the occurrence of the field induced ordering.
At this temperature the data for $x~=~0$ can be clearly fit with two peaks, see panel (a), and these two peaks correspond to those observed in Fig.~\ref{f112}a for T~=~300~K and $H \geq 7$~T.
\begin{figure}[b]
\centerline{
\epsfig{figure=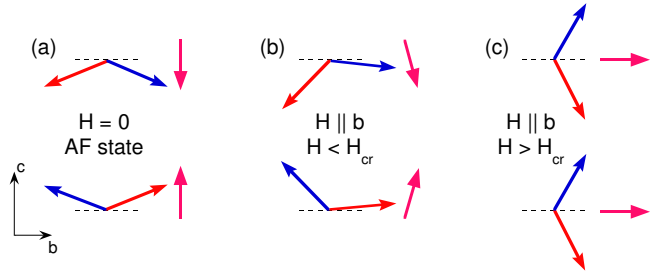,width=90mm}
}
\caption{
Cartoon showing the proposed changes in the spin structure in x~=~0 and 0.01 with the application of magnetic fields $\vec{H} \parallel \hat{b}$ axis, starting from the AF state, panel (a).
For small fields there is a slight rotation of the WF moments, panel (b), which will lie in the $(ab)$ planes at high fields, panel (c). 
}
\label{f114}
\end{figure}

The temperature dependence of the FIM across the N\'{e}el boundary can be studied in the x~=~0.01 crystal which has a lower T$_{N}$, see the panels (b) and (d) from Fig.~\ref{f113}.
We mention here that for the $x~=~0.01$ \lsco crystal T$_{N}$ was measured (for fields lower than 7~T) to decrease on the average by almost 4~K/T for fields $\vec{H} \parallel \hat{b}$ axis.
Given a $T_{N} (0 T) \approx 215$~K, in a 9~T field one expects that $T_{N} (9~T) \approx 180$~K.  
Indeed, at 9~T and below 180~K the behavior for x~=~0.01 is very similar to that in the undoped crystal showing a softening of the FIM as we warm to T$_{N}$ but the situation changes with further warming.
The 200~K data show that the FIM has, similarly to the 295~K data for x~=~0, a low energy shoulder which is marked by an arrow.
The data at 250 and 275~K can also be fitted by two Lorentzians.
Along with the 200~K spectrum, these data seem to suggest the following picture: above 180~K we observe two features, one whose energy does not show significant magnetic field dependence and another one which softens from 20 to about 8~\cm-1 with decreasing the temperature from 275 to 200~K.
This latter excitation is marked by arrows in Fig.~\ref{f113}c and its energy is plotted in panel (d) by empty circles. 
The filled red circles in the same panel show the energy of the other mode (whose frequency is almost magnetic field independent above 180~K).
The solid lines offer an explanation for the softening of the peak marked with arrows in panel (b): this is a magnetic soft mode corresponding to the field induce spin order taking place at $T \approx 180$~K in x~=~0.01 \lsco for $\vec{H} \parallel \hat{b}$ and $H = 9$~T.
Its energy approaches zero on cooling towards T$_{N}$ and we propose that it becomes the DM gap in the N\'{e}el phase.
Fig.~\ref{f113}d also shows that the plot of the integrated intensities of the FIMs as a function of temperature is peaked at T$_{N}$.
This points toward an unusual behavior in the sense that in a conventional picture the intensities of long wavelength gap modes scale with the AF order parameter, i.e. both of them vanish as T$_{N}$ is approached from below~\cite{KeimerZP93}.

What is the nature of the FIM within the AF phase?
A possible explanation is its identification to the $XY$ gap.
Support for this assignment is the presence of this mode only in x~=~0 and 0.01 \lsco as well as the comparison to INS data~\cite{KeimerPRB92,KeimerZP93} which estimates $\Delta_{XY} \approx$~40~\cm-1 at 10~K in \lco.
The very small experimentally found hardening of the FIM with increasing field from 4.5 to 9~T at T~=~10~K shown in Fig.~\ref{f112}e is consistent within 20\% with the predictions of the spin-wave theory, which was found to describe fairly well the DM gap.
This difference may be also accounted for if one invokes possible gap renormalization effects induced by higher order spin interactions \cite{ColdeaPRL01}.

Regarding the nature of the magnetic field induced order we propose a state like the one depicted in Fig.~\ref{f114}c.
This is suggested by the magnetic susceptibility data which shows that the moments on Cu sites remain confined in the $(bc)$ plane above T$_{N}$~\cite{LavrovPRL01} and also by recent magnetoresistance
measurements~\cite{AndoPRL03} which are consistent with a gradual rotation of the WF moments.
In fact a departure from a two step transition \cite{ThioPRB90}, involving a spin-flop process occurring between the states shown in Fig.~\ref{f114}b-c and which is characterized by a large component of the staggered magnetization along the $a$ orthorhombic axis, is expected.
In a regular spin-flop transition  a magnetic field applied parallel to the easy axis (which in our case is the $b$ orthorhombic axis, see Fig.~\ref{f16}b) will end up rotating the staggered magnetization along a direction perpendicular to this axis.
The reason is that above some critical value of the field the magnetic anisotropy energy becomes smaller than the gain in magnetic energy due to the larger transverse susceptibility in the AF state \cite{KefferBook}.
In the \lsco case, the situation seems not to be the same: because the transverse susceptibility, $\chi_{a}$, is the smallest below 300~K for x~=~0 and 0.01 (see Fig.~\ref{f13}), the spins cannot partake of the field energy $- (\chi_{a} - \chi_{b}) H^{2} / 2$.
Accordingly, a flop along the $a$-axis is not favorable from this point of view \cite{Ono04}.

The identification of the FIM with the $XY$ gap can also explain other observed features.
The crossover around 150~K shown in Figs.~\ref{f112} and \ref{f113} may be understood as a departure of the direction of the WF moments from perpendicular to the $(ab)$ plane to a direction almost parallel to
the $b$-axis (see Fig.~\ref{f114}) where the $XY$ anisotropy, weaker due to temperature fluctuations, ceases to play a decisive role.
Physically,  this corresponds to the fact that the conventional out-of-plane XY mode changes its nature as the WF moment rotates away from the $c$-axis.
Prompted by this idea we calculated (solid lines in Fig.~\ref{f112}c) the spin-wave dispersions using Eq.~(\ref{e15}) and (\ref{e16}) in the extreme case of $\alpha = 0$ and a small DM gap which still confines the moments in the $(bc)$ plane.
Although finite temperature effects have to be taken into account, we note that this simple estimation reproduces, at least qualitatively, the experimental dispersions.
We also comment on the possible relevance of our findings to the switch of orthorhombic axes in magnetic fields~\cite{LavrovNature02}.
If a state like Fig.~\ref{f114}c is realized (which is shown in Fig.~\ref{f113}b to persist to temperatures close to 300~K even for x~=~0.01) then the magnetic force in an external field is significantly enhanced due to the net in-plane ferromagnetic moment.
Still, the origin of the coupling between the spins and the tilt of the CuO$_{6}$ octahedra remains as a very interesting question.

The qualitative scenario we propose regarding the nature of the FIMs and the nature of the magnetic field induced order leaves several open questions.
One of them is the following: if the FIM in the AF state is the XY gap, why is its spectral weight peaked at T$_{N}$, as shown in Fig.~\ref{f114}b for x~=~0.01 \lsco?
A second question is related to the finite intensity of the FIMs only for magnetic fields $\vec{H} \parallel \hat{b}$-axis.
On the other hand if we assume that the FIM mode is an excitation other than the XY gap, arising for instance as a result of the 4-sublattice structure, then the common interpretation of the excitation around 40~\cm-1 found in several 2D layered AF's has to be reconsidered.

One may wonder if up to now there are any other transport signatures of this magnetic field induced transition which could back up our spectroscopic conclusions.
As for the undoped \lco crystal, where the transition should be most prominent and which has strongly insulating behavior, to our knowledge there are no magnetoresistance measurements so far and in terms of magnetization it would be highly desirable to see measurements especially as a function of magnetic field at several temperatures down to 10~K.
Higher fields than 7~T are needed though as the temperature is decreased below 300~K.
However, relative magnetoresistance data in twinned samples of x~=~0.01 \lsco (Fig.~2 in Ref.~\cite{AndoPRL03}) show a 'peel off' from a temperature independent curve.
For a given magnetic field value, this phenomenon is seen to occur at temperatures where magnetization data indicate the transition outside the AF order.
This shows that the $dc$ transport responds to field induced changes in the AF environment in the anticipated (H,T) parameter space.
Supplementary magnetization, magnetoresistance and especially neutron scattering measurements in magnetic field would be necessary to verify our claims.
\begin{figure}[t]
\centerline{
\epsfig{figure=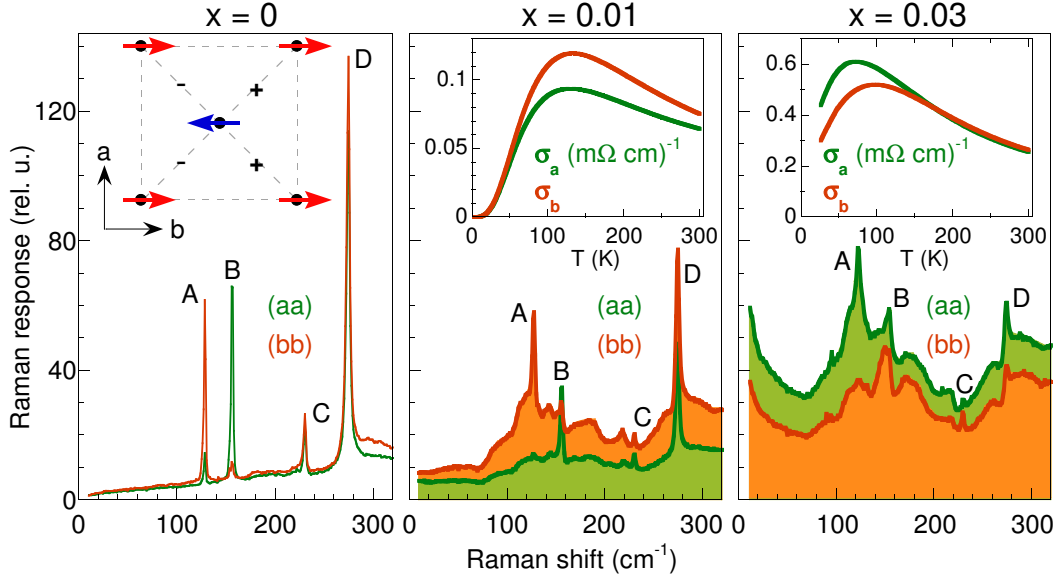,width=140mm}
}
\caption{
Main panels show (from left to right) T~=~10~K Raman data in $(aa)$ and $(bb)$ polarizations in x~=~0, 0.01 and 0.03 \lsco.
$A$, $B$, $C$ and $D$ denote 4 of the 5 fully symmetric Raman modes in the LTO phase. 
The inset in the left panel shows a unit cell of the LTO structure, the directions of spins on Cu sites and the axes notations (see also Figs.~\ref{f12} and \ref{f16}).
The insets for the x~=~0.01 and 0.03 panels show $dc$ conductivities along the $a$ and $b$ axes.
The same vertical scale was used for the Raman spectra in panels (a), (b) and (c). 
}
\label{f115}
\end{figure}

\subsection{Phononic and Electronic Anisotropy in Detwinned  \lsco}
We observed in the previous section that detwinned \lsco crystals revealed strong anisotropy effects in terms of the dynamics of long wavelength spin excitations.
Here we show that the small lattice orthorhombicity has drastic effects also on the phononic and electronic Raman continuum.
A summary of our results in this perspective is shown in Fig.~\ref{f115}.
The left panel shows T~=~10~K Raman data in \lco taken in $(aa)$ and $(bb)$ polarizations.
The axes notation is shown in the inset.
Both these symmetries probe fully symmetric excitations and, in terms of phonons, there are 5 allowed in the LTO phase.
Four of them, denoted by $A$, $B$, $C$ and $D$ are in the energy region below 300~\cm-1.
All the five A$_{g}$ modes and their atomic displacements will be discussed in more detail in the section devoted to Nd doped \lsco.
For now we remark that while in the insulating \lco the Raman continuum is, as expected, very weak at low temperatures, there is a tremendous intensity anisotropy in the $A$ and $B$ phonons.
Mode $A$ is seen clearly in $(bb)$ polarization and has almost vanishing intensity in $(aa)$ polarization and the situation is reversed for mode B.  

In the middle panel we observe that the anisotropy in these two modes is preserved.
However one can note that we observe intense Raman backgrounds (shaded areas), quite different in intensity in $(aa)$ versus $(bb)$ polarizations.
The relative intensities of the continua match the anisotropy in the  $dc$ conductivity along the $a$ and $b$ axes \cite{AndoPRL02} shown in the inset.
Looking at the $x~=~0.03$ data (right panel) one can notice that the sign of the low temperature resistivity anisotropy changes with respect to the $x~=~0.01$ case.
Similarly, the Raman background in $(aa)$ polarization becomes stronger than in $(bb)$ polarization and this change is also accompanied by the reversal of the intensity anisotropy of the $A$ and $B$ phonons.

This switch is a remarkable effect.
Could it be that it is induced by structural changes, in particular a 90$^{\circ}$ rotation of the CuO$_{6}$ octahedra between 1 and 3\% Sr doping?
X-ray data showed that this is not the case, suggesting that the reversal is due to the development of a new kind of anisotropy in the spin-charge dynamics at low doping, possibly occurring as the the system crosses at low temperatures the boundary of the long range AF order.
Beyond this observed switch between 1 and 3\% doping, the strong phononic anisotropy seen most clearly in \lco data is an interesting problem by itself.
Since it is determined by the CuO$_{6}$ octahedra tilt around the $a$-axis, one may suspect that the $p_{z}$ orbitals of the apical oxygens may be involved in the coupling process and its hybridization with in plane orbitals is not negligible.
We remark one other point in regard to the phononic features shown in Fig.~\ref{f115}: While in \lco the observed number of modes does not exceed the number predicted by group theory (we observe also the 5$^{th}$ mode in $(cc)$ polarization at 430~\cm-1, see Fig~\ref{f120}), a much larger number of additional features sitting on top of the Raman continuum is seen for $x~=~0.01$ and 0.03.
It is possible that this is connected to charge and/or spin supermodulation within the 2D CuO$_{2}$ planes.
While not explained, the experimental observations in Fig.~\ref{f115} pose intriguing questions, some of them, like the possibility of 2D spin and/or charge order, being tied to problems actively scrutinized in relation to the occurrence of superconductivity in cuprates.

\section{Spin and Lattice Dynamics at Commensurate $x~=~1/8$ Sr  \newline Doping in
\lnsco}

\subsection{Motivation: Intrinsic Spin/Charge Modulations in the CuO$_{2}$ planes?}

The origin of the interest in studying lattice and electronic dynamics in 2D cuprates at carrier concentrations commensurate with the lattice is essentially due to the increased tendency of the doped system to form real space patterns characterized by certain periodic modulations of the charge and spin density.
Among correlated systems, this situation is not peculiar to high T$_{c}$'s but it has been discussed for instance in different type of materials like manganites or nickelates.
Ground states in which charges self organize in quasi-1D 'rivers' (called stripes) acting as AF domain walls were predicted at the mean field level as early as 1989 \cite{ZaanenPRB89} and later it has been proposed that the charge and/or spin ordering is not necessarily static, but the carriers could form electronic liquid-crystal like phases \cite{KivelsonNature98}.

\begin{figure}[t]
\centerline{
\epsfig{figure=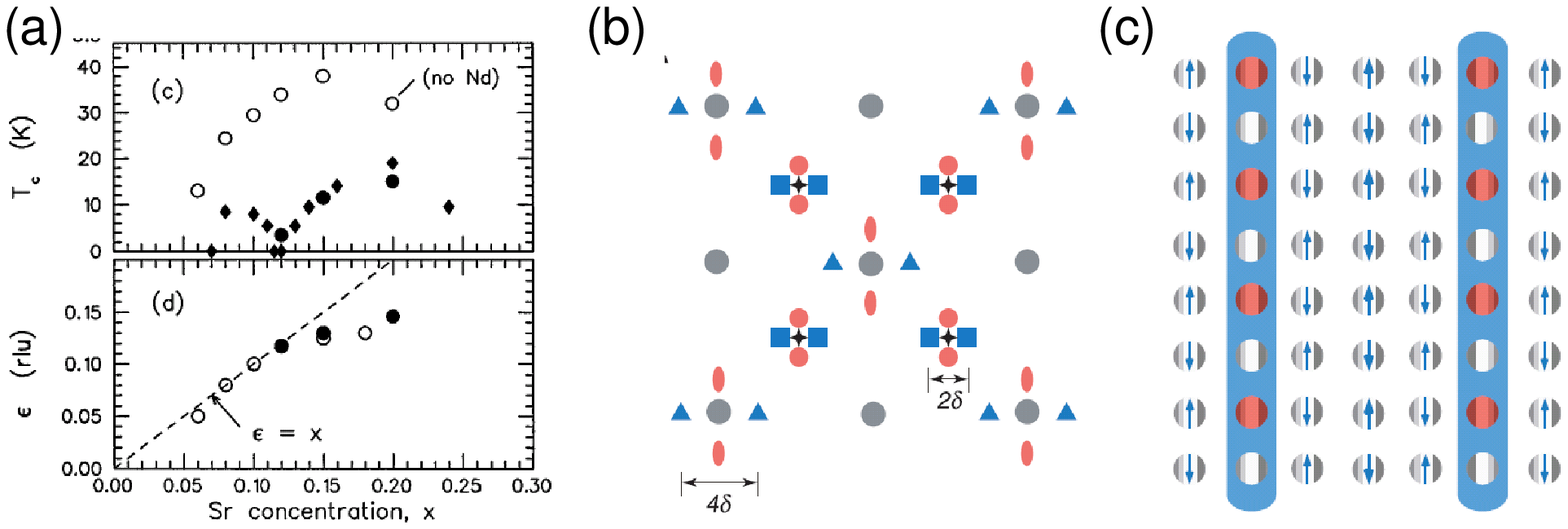,width=130mm}
}
\caption{
(a) The superconducting temperatures (upper panel) and the evolution of the incommensurate splitting (lower panel) in \lsco (empty symbols) and $y = 0.4$ \lnsco (filled symbols) as a function of Sr concentration $x$.
Data from Ref.~\cite{TranquadaPRL97}.
The symbol $\epsilon$ in (a) is denoted by $\delta$ in (b) and (c).
Neutron scattering: reciprocal and real space are shown in panels (b) and (c).
In (b) the grey circles at $(2 \pi / a) (m,n)$ with $(m,n)$ being a pair of integers correspond to fundamental structural Bragg peaks in the HTT phase.
The black stars at $(2 \pi / a) (m,n) + (\pi / a, \pi / a)$ are for the commensurate AF Bragg peaks.
The points separated by $2 \delta$ are incommensurate dynamic or static peaks which replace the AF Bragg peaks upon Sr doping.
Their associated wavelength is $\lambda_{spin} = 1 / \delta$.
The points separated by $4 \delta$ correspond to the charge order with $\lambda_{charge} = 1 / 2 \delta$ in $y = 0.4$ \lns0418.
In (c) the real space charge-spin structure inferred from the results in (b).
Charges are confined in the blue channels and the blue arrows on the external grey circles indicate the magnitude and direction of the magnetic moments on Cu sites.
}
\label{f116}
\end{figure}

From the experimental point of view, one of the observed '$x~=~1/8$' effects, discovered initially in \lbco \cite{MoodenbaughPRB88} but also observed in Nd doped \lsco \cite{TranquadaPRL97}, was a suppression of superconductivity manifested through a decrease of the transition temperature T$_{c}$.
In fact a similar observation (but in terms of the \emph{onset} of superconductivity as seen by magnetization measurements \cite{YamadaPRB98}) was made in Nd free \lsco at 1/8 Sr doping.
The 'stripology' in cuprates got a lot of momentum after the discovery of a constellation of neutron Bragg peaks in \lns0418.
The data showed superlattice peaks associated with static spin and hole ordering, the magnetic moment modulation being characterized by a wavelength  twice as big as the one observed for the charge \cite{TranquadaNature95}.
Some of the effects discussed above are illustrated in Fig.~\ref{f116}.

The almost complete suppression of T$_{c}$ in \lns0418 as well as the fact that in this compound neutron scattering sees long ranged charge and spin supermodulations suggested that the stripes may be the 'looked for' competing state to superconductivity.
The presence of the incommensurate magnetic peaks also in Nd free \lsco and their observation in \emph{elastic} neutron scans at $x~=~1/8$ doping show that \lnsco with $x~=~1/8$ are some of the most suitable 2D cuprate compounds to look for the effects of such modulations.
It is important to note that while in Nd doped \lsco the charge ordering was also confirmed by X-rays \cite{ZimmermannEL98}, this is not the case (yet) in Nd free samples.
Raman spectroscopy can be a powerful technique in this respect because optical phonons can be used as local probes of fast changes in the charge distribution and magnetic Raman scattering provides information about local AF correlations.
However, it can also provide information regarding the side effects of Sr substitution and what we argue in the study presented in the following is that some of those effects, structural distortions as well as the disorder introduced by Sr substitution, are important at $1 / 8$ doping in \lsco irrespective of Nd concentration \cite{GozarPRB03}.

\begin{figure}[b]
\centerline{
\epsfig{figure=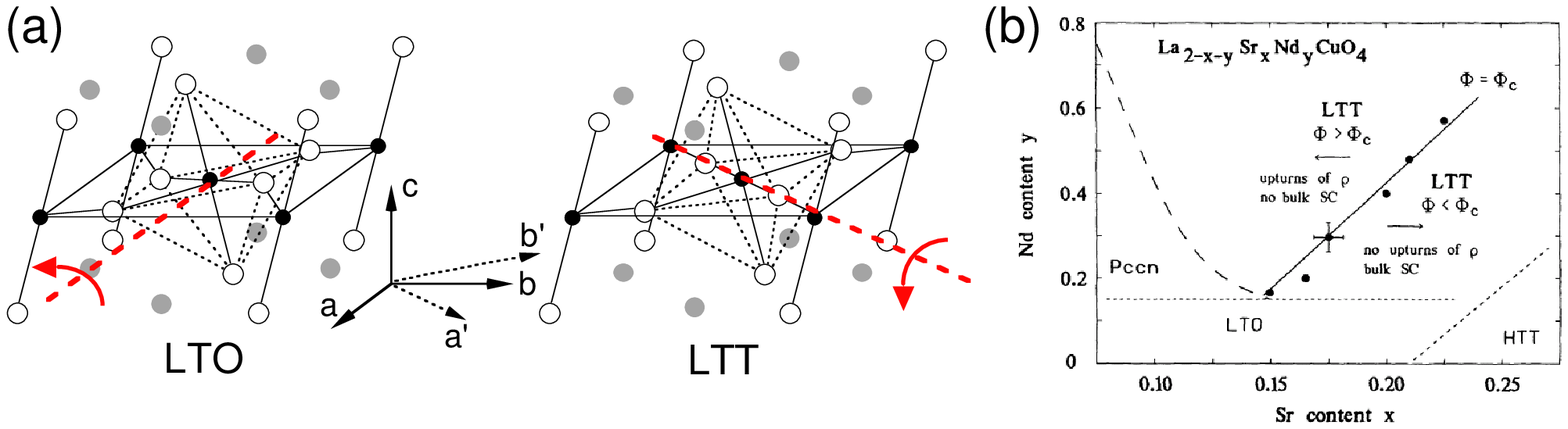,width=150mm}
}
\caption{
(a) Buckling of the CuO$_{2}$ planes and the tilt pattern of the CuO$_{6}$ octahedra in the LTO and LTT phases.
(b) A schematic of the T~=~10~K structural phase diagram of \lnsco as a function of x(Sr) and y(Nd) (from Ref.~\cite{BuchnerPRL94}).
The main point of this plot is that the LTT region characterized by a CuO$_{6}$ tilt angle $\Phi > \Phi_{c} \approx 3.6^{\circ}$ has no bulk superconductivity (SC) and low temperature insulating behavior while SC is found in the LTT region with $\Phi < \Phi_{c}$.
In other words it is proposed that there is a critical buckling angle compatible with long range SC and that the stabilization of the LTT phase suppresses SC correlations.
}
\label{f117}
\end{figure}

It was mentioned in the introduction that in \lnsco the various changes in the crystal structure are due to the lattice mismatch between the cation and CuO$_{2}$ layers.
Like \lsco, the \lns0418 compound undergoes a transition from the HTT to the LTO phase above room temperature.
This transition is followed around T$_{LTT} = 70$~K by another structural change, from the LTO to the low temperature tetragonal (LTT) phase where the CuO$_{6}$ octahedra tilt around an axis parallel to the Cu-O-Cu bonds.
The structural order parameter of these transitions is the libration of the CuO$_{6}$ octahedra shown in Fig.~\ref{f117}.
It was noticed in \lns0418 that the LTO-LTT transition takes place over a range of temperatures and that disorder in the striped phase leads to a glassy nature of the ground state.
Intermediate states characterized by a tilt angles in between those of the LTO and LTT phase have also been proposed \cite{BuchnerPRL94}.
The coexistence in \lnsco of several phases in a complex mixture was suggested by transmission electron microscopy~\cite{HoribePRB00}.

So there are interesting topics associated to the presence of Nd, but after all why are these structural effects, especially the ones related to the tilt of CuO$_{6}$ octahedra, relevant to the spin and charge dynamics?
The importance of the local structural distortions for the superconducting properties characterized in cuprates by a short coherence length should not be ignored.
The stabilization of the LTT phase was observed to trace the suppression of superconductivity in Nd doped \lsco \cite{CrawfordPRB91} and also in the related La$_{2-x}$Ba$_{x}$CuO$_{4}$ compound \cite{AxePRL89}.
A critical value of the CuO$_{6}$ tilt was associated with the stabilization of magnetic against superconducting order, see Fig.~\ref{f117}b and Ref.~\cite{BuchnerPRL94}.
Rapid suppression of superconductivity, similar to that due to Cu replacement by non-magnetic impurities, was observed with increasing the cation radius variance~\cite{McAllisterPRL99}.

In this context our study provides direct spectroscopic information about the LTO-LTT transition in \lns0418 and local deviations from the average structure existent in Nd doped and Nd free \lsco structures.
The persistent fluctuations of the structural order parameter down to T~=~10~K reveal substantial disorder in the cation-oxygen layers.
The distinct Raman signatures accompanying a transition to a state with deep spin/charge modulations are not observed in the temperature dependence of the two-magnon (2M) scattering around 2200~\cm-1 and the $c$-axis polarized phonons below 500~\cm-1 \cite{GozarPRB03}. 

\subsection{Inhomogeneous CuO$_{6}$ Octahedra Distribution in $x = 1/8$ \lnsco}

In the following we will show Raman data from \lnsco with the following doping concentrations: $x \approx 1/8$, $y = 0$;  $x \approx 1/8$, $y = 0.4$; and $x = 0.01$, $y = 0$.
The spectra were taken from the $(a'c)$ and $(ab)$ faces of the $x = 1/8$ \lnsco samples and from the $(ac)$ surface of a $x = 0.01$ \lsco crystal as determined by X-ray diffraction.
See Fig.~\ref{f117} for axes notations. 
Note that they are consistent with the ones in Fig.~\ref{f12}, the primed letters corresponding to directions parallel to the Cu-O-Cu bonds
The laser excitation energy used was $\omega_{in} = 1.92$~eV.

\begin{figure}[b]
\centerline{
\epsfig{figure=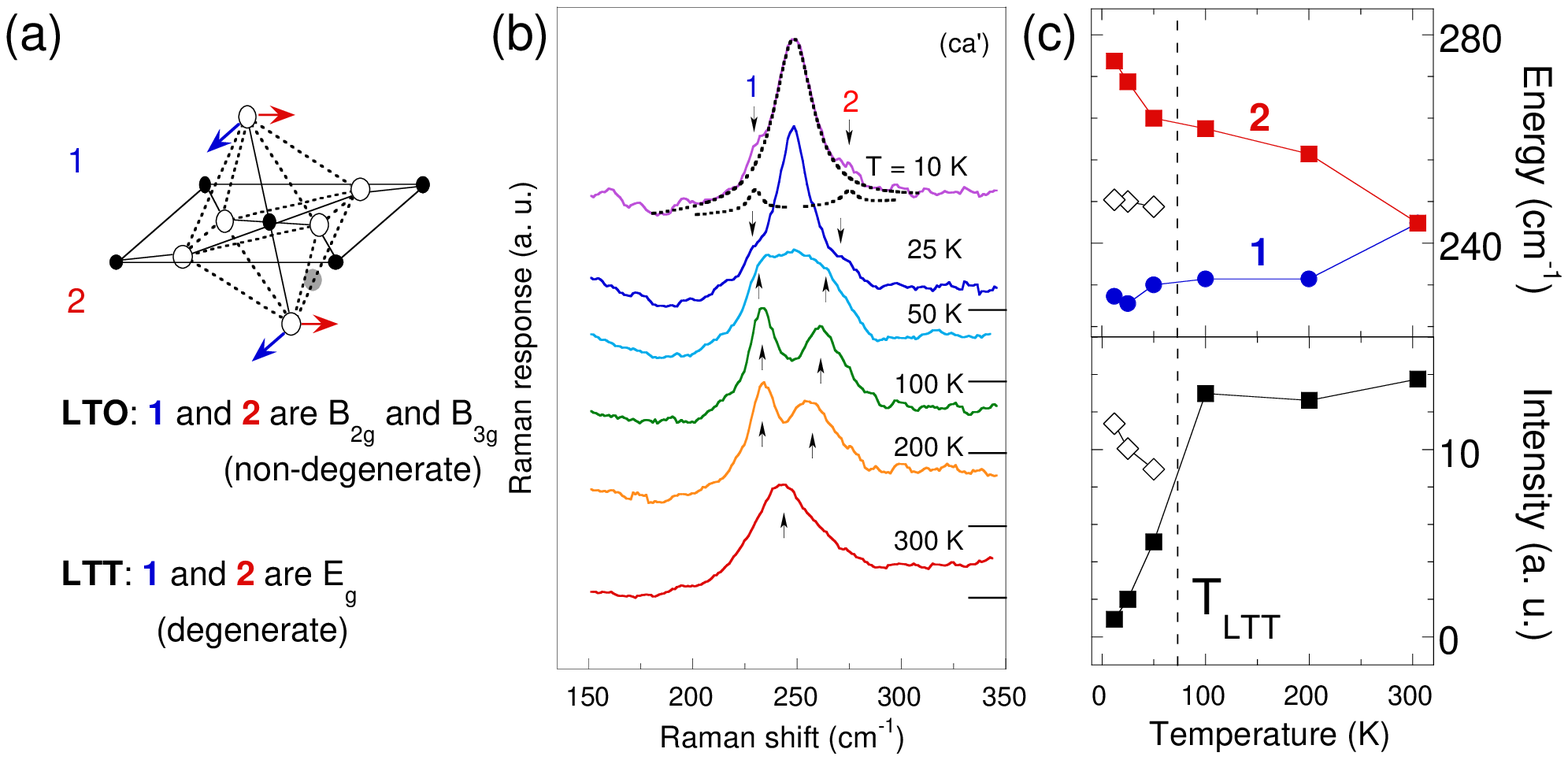,width=130mm}
}
\caption{
(a) Cartoon with the apical Oxygen vibrations having B$_{2g}$ and B$_{3g}$ symmetry in the LTO phase and degenerate with E$_{g}$ symmetry in the LTT phase.
(b) Temperature dependence of the O modes described in (a) in \lns0418.
The finite intensity of modes 1 and 2 even at T~=~10~K shows that there exists a residual orthorhombicity at this temperature.
(c) The temperature dependence of the energies (upper panel) and intensities (lower panel) of the phonons in (b) obtained by Lorentzian fits.
}
\label{f118}
\end{figure}
Raman spectra taken in the $(ca')$ geometry may provide direct information about tetragonal to orthorhombic distortions.
In this polarization we probe phononic modes with B$_{2g}$ and B$_{3g}$ symmetries in the LTO phase which become degenerate with E$_{g}$ symmetry in the LTT phase.
In Fig.~\ref{f118} we show the temperature dependence of the modes around 250~\cm-1 corresponding to the apical O vibrations parallel to the CuO$_{2}$ plane in \lns0418 \cite{OhanaPRB89}.
One can think about the spectral changes in analogy to the evolution with temperature of the
orthorhombically split X-ray diffraction Bragg peaks \cite{CrawfordPRB91}.
We observe a broad peak around 245~\cm-1 at room temperature which, with cooling, becomes resolved into two components, one hardening and one softening.
A new central peak can be seen at 50~K around 248~\cm-1 which gains spectral weight as the temperature is decreased to 10~K.
While the total integrated intensity of the modes remains constant, Fig.\ref{f118}c, we observe a redistribution of spectral weight among the three modes as a function of temperature.
The split components become weaker but can still be seen as 'orthorhombic satellites' of the central peak down to 10~K.
The coalescence of the features into the 248~\cm-1 mode signals the recurrence of a phase with tetragonal symmetry which should be the expected LTT phase of \lns0418.
However, the finite residual intensity of the satellites appearing on the tails of the broad central peak shows an incompletely developed LTT phase and that even at 10~K there exists about 7\% LTO phase 
(determined from the relative ratio of phonon intensities).
Note that the width of the main peak at T~=~10~K is comparable to the widths of the components of the doublet seen at temperatures as high as 200~K.

\begin{figure}[t]
\centerline{
\epsfig{figure=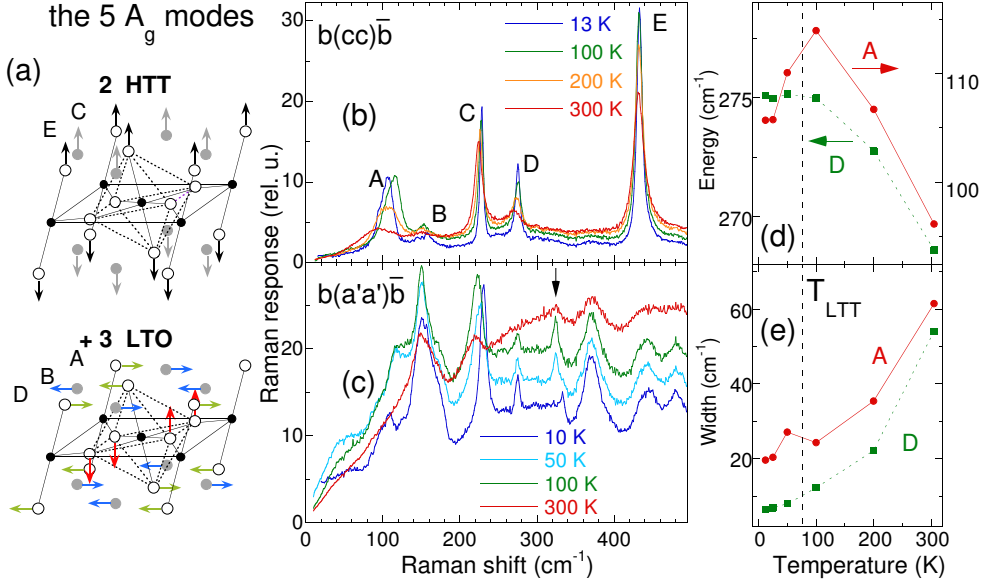,width=130mm}
}
\caption{
(a) Group theory predicts two A$_{1g}$ modes in the HTT phase (denoted by C and E) and the appearance of 3 additional A$_{g}$ phonons in the LTO phase (denoted by A, B and D).
They are linear combinations of the atomic displacements shown in this panel.
Temperature dependence of the $b(cc){\bar b}$ (panel b) and $b(a'a'){\bar b}$ (panel c) polarized Raman spectra.
In (d) and (e) we show the temperature dependent energies and intensities of modes A and D from panel (b).
}
\label{f119}
\end{figure}

Raman data in $(cc)$ polarization is well suited for the study of lattice dynamics due to weaker coupling to underlying electronic excitations.
Temperature dependent Raman spectra in this scattering geometry are shown in Fig.~\ref{f119}b. 
Group theory predicts five fully symmetric modes at $k = 0$ in each of the LTO and LTT phases and only two for the HTT phase.
The two fully symmetric modes of the HTT along with the additional three modes in the LTO phase are shown in Fig.~\ref{f119}a.
In Fig.~\ref{f119}b we observe all the five phonons corresponding to the LTO and LTT phases and they are denoted by A, B, C, D and E.
Four out these five modes can also be seen in Fig.~\ref{f115} where the same notation was used.
Although every one of these excitations should be considered as linear combinations of all the A$_{g}$ movements depicted in Fig.~\ref{f119}a, one could roughly say that they are mainly composed of the following vibrations.
The modes C and E (which at T~=~10~K are found at 228 and 433~\cm-1) are inherited from the HTT phase and they correspond to the $c$-axis vibrations of La/Sr/Nd and O atoms respectively, see the upper part of Fig.~\ref{f119}a.
Mode A is the soft mode of the HTT-LTO transition (the CuO$_{6}$ octahedra tilt), mode B is mainly due to  the vibration of La/Sr/Nd atoms in the direction imposed by the CuO$_{6}$ tilt and mode D consists of $c$-axis vibrations of the in-plane O atoms \cite{OhanaPRB89,SugaiPRB89,WeberPRB88}.
The energies of the last three phonons at the lowest temperature are: 106~\cm-1 (mode A), 156~\cm-1 (mode B) and 275~\cm-1 (mode D).

\begin{figure}[t]
\centerline{
\epsfig{figure=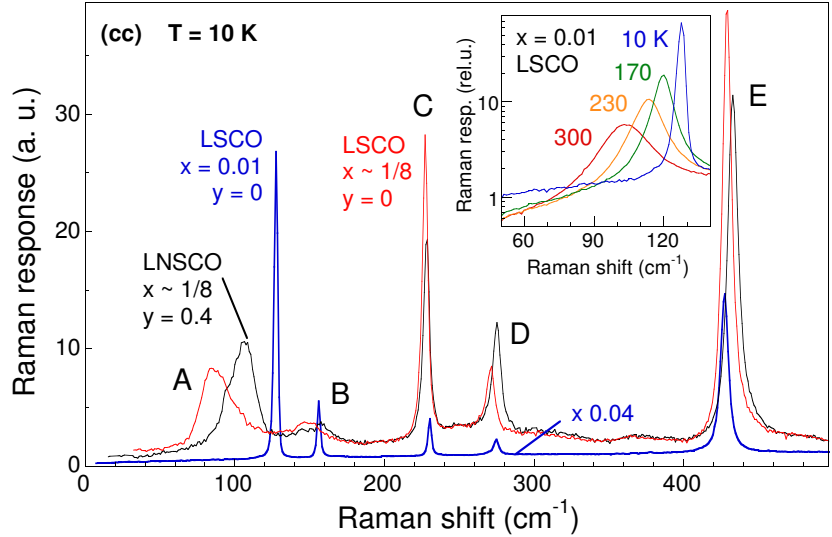,width=110mm}
}
\caption{
T~=~10~K Raman spectra in $b(cc){\bar b}$ configuration for $x = 1/8; y = 0$ (red), $x = 1/8; y = 0.4$ (black) and $x = 0.01; y = 0$ (blue) \lnsco.
The inset shows the temperature dependence in $b(cc){\bar b}$ polarization of the the intensity of mode A (CuO$_{6}$ octahedra tilt) in x~=~0.01 \lsco.
}
\label{f120}
\end{figure}

The above qualitative description indicates that we could expect a strong coupling between the lowest energy modes (A and B).
These two excitations can be distinguished in Fig.~\ref{f119}b from the other ones because they remain much broader and look like composite features even at the lowest temperature in comparison with the the modes C, D and E which harden and sharpen smoothly through the LTO-LTT transition taking place around 70~K.
As seen in Fig.~\ref{f119}b, the temperature variation of the intensities of the modes C and E inherited from the HTT  phase is not as pronounced which is not surprising.
A comparison of the temperature dependent energy and full width at half maximum (FWHM) of modes A and D is shown in Fig.~\ref{f119}d-e.
The large variation in energy and width of mode A above the transition (see also the inset of Fig.~\ref{f120}), the softening below 70~K, as well as its energy around 110~\cm-1 in agreement with neutron scattering studies~\cite{ThurstonPRB89} show that this mode corresponding to the octahedra tilt is the soft mode of the structural changes \cite{SugaiPRB89}.
The smooth decrease in the energy in the LTT phase is only apparent because this space group is not a subgroup of the LTO group and as a result a true LTO-LTT transition is expected to be of first order.
Although unresolved due to broadening effects, the large width of mode A around 70~K shows the coexistence of the LTO and LTT tilts, the latter appearing as a result of folding of the LTO $Z$-point to the $\Gamma$-point of the LTT phase which was observed also in La$_{2}$NiO$_{4}$~\cite{BurnsPRB90}.

We infer from our data that the large FWHM of mode A reflects the spatial distribution of the octahedra tilt.
The simultaneous broadening of the mode B shows coupling between the OP and La/Nd/Sr vibrations and as a result the influence of the dynamics in the cation-O layers on the properties of CuO$_{2}$ planes.
Both modes B and C involve cation displacements as discussed above, the former perpendicular and the latter parallel to the $c$ axis.
However, at 10~K the FWHM of mode C is 8~\cm-1, smaller compared to the FWHM of mode B which is around 20~\cm-1.
We conclude that the large observed widths of the modes A and B are mainly caused by the locally fluctuating OP and not due to the inhomogeneous broadening introduced by the simultaneous presence of La, Nd and Sr in the inter-layer composition which should have been reflected also in a large width of mode C.

It is interesting to compare the $(cc)$ polarized phononic spectra with those in which the polarization of the incoming photon field is parallel to the CuO$_{2}$ planes.
Fig.~\ref{f119}c shows the temperature dependent Raman spectra in the $b(aa)\bar{b}$ geometry.
Different coupling to the electronic degrees of freedom when the polarization of the incident field is parallel to the CuO$_{2}$ planes leads to a stronger intensity in the underlying Raman continuum relative to the phononic features and also a different intensity/shape of the fully symmetric features observed in $(cc)$ polarization.
A continuous suppression with cooling of the electronic background is due to the opening of a pseudogap in \lns0418 \cite{DummPRL02}.
The electron-phonon coupling allows the observation of additional peaks around 370 and 480~\cm-1 evolving smoothly from 300 to 10~K, both of which allowing an interpretation in terms of two-phonon scattering if some anharmonic interaction is taken into account.
Marked by an arrow in this panel is a B$_{1g}$ symmetric excitation in the LTO phase which shows a jump from 325 to 335~\cm-1 as the crystal enters in the LTT phase.

In order to understand the surprising behavior of the tilt pattern as reflected in the phononic data from Fig.~\ref{f119} a comparison with different materials from the same class is useful.
In Fig.~\ref{f120} we show the 10~K $(cc)$ polarized Raman spectra of three crystals: \lns0418, x~=~0.01 and 0.12 \lsco.
For $x = 0.01$ \lsco mode A has a FWHM of 2.5~\cm-1 (Note in the inset the strongly temperature dependent intensity and width which is a characteristic of a soft mode).
For $x \approx 1/8$~LSCO the same phonon is around 85~\cm-1 and its FWHM of about 23.5~\cm-1 is larger than the width of mode A in the Nd doped crystal where it is slightly below 20~\cm-1.
Comparison of these relative phononic widths confirms the conclusion discussed before that Nd doping of LSCO crystals and the closer proximity to the T$^{\prime}$ phase induced by Nd doping in the La$_{2}$CuO$_{4}$ structure \cite{ManthiramJSSC91} cannot be responsible for the large observed broadening effects.
Intrinsic phonon anharmonicity would lead to a broad mode A in $x = 0.01$~LSCO which is not the case.
Neither can the tilt disorder across twin domains be the cause of such dramatic effects because the volume fraction occupied by these boundaries is expected to be very small \cite{HoribePRB00}.
The 7\% relative ratio of the orthorhombic satellites to the central peak in Fig.~\ref{f118}b would rather be consistent with such a small contribution.
If the satellites are indeed due to twinning effects the data show that at 10~K the larger LTT domains are separated by regions of pure LTO tilt.
The absence of the broadening effects on the vibrations along the $c$-axis points towards an 'anisotropic' disorder relating primarily to bond randomness along directions parallel to the CuO$_{2}$ planes.

Could the spin-lattice coupling or the interaction with the stripe-ordered carriers in CuO$_{2}$ planes be the main cause of broadening?
Stripe correlations are enhanced in \lns0418 which displays however a smaller width of mode A.
Also, it is not clear why only the modes A and B would be affected by this interaction.
In this sense one expects the movements of the in-plane atoms to be more sensitive to stripe ordering but we see no similar effects on mode D.
Although less probable, spin-lattice induced broadening cannot be completely ruled out and the answer to this question lies in a Sr doping dependence of the $(cc)$ polarized spectra.
Our data can be reconciled however with recent studies of local structure in Nd free and Nd doped \lsco systems \cite{HaskelPRL96HanPRB02}.
Model analysis of the pair distribution function from X-ray absorption fine structure suggests that in this material class the average structure determined by diffraction is different from the local pattern
which is characterized by disorder in the CuO$_{6}$ tilt direction and magnitude~\cite{HaskelPRL96HanPRB02}.
The Raman data shown in Figs.~\ref{f119} and \ref{f120} are spectroscopic evidences that the \lnsco system is characterized by disorder in the cation layers and that the locally fluctuating octahedra tilt is responsible for the observed effects.

Information about the relative magnitude of charge disproportionation in \lnsco can be gained by comparison with Raman spectra in compounds like the nickelates \cite{BlumbergPRL98,PashkevichPRL00} or manganites \cite{AbrashevPRB01,YoonPRL00} where charge and spin modulations are well established \cite{ChenPRL93ChenPRL96}.
New Raman active modes have been observed below the charge ordering within the Mn-O layers in La$_{0.5}$Ca$_{0.5}$MnO$_{3}$ \cite{AbrashevPRB01} and also in
Bi$_{1-x}$Ca$_{x}$MnO$_{3}$ \cite{YoonPRL00}.
Conspicuous changes in the lattice dynamics have also been observed in x~=~0.33 and 0.225 La$_{2-x}$Sr$_{x}$NiO$_{4}$ by Raman scattering \cite{BlumbergPRL98,PashkevichPRL00}.
Lowering of the crystal symmetry at the stripe ordering transition
gives rise to folding of the Brillouin zone and the appearance of new $k = 0$ phononic modes.
Charge localization creates non-equivalent Ni sites generating phonon 'splitting'.
The $c$ axis stretching modes corresponding to La and apical oxygens split by 14 and 30~\cm-1 respectively \cite{BlumbergPRL98}.
Within about 3~\cm-1 resolution imposed by the phononic widths we do not observe such splittings in our spectra.
The ratio of the integrated intensities of the split oxygen modes in Ref.~\cite{BlumbergPRL98} is about the same as the ratio of doped versus undoped Ni sites.
If we assume the same relation to hold for the case of cuprates, a factor of 12\% in split phononic intensity should have been seen in our spectra.
However, the latter argument has to take into account that different electron-phonon coupling might change this proportionality relation.
Last but not least is the observation that we see, at least in $(cc)$ polarized spectra only the phononic excitations predicted by group theory for the LTO/LTT phases. 
We conclude that any charge ordering taking place in our case is much weaker than in the related compounds referred to above.
This is not contradicting X-ray diffraction data \cite{ZimmermannEL98} which estimated a factor of 10$^{2}$ between the relative magnitude of charge modulations in cuprates and nickelates.

\subsection{Two-Magnon Raman Scattering in $x = 1/8$ \lnsco and $x = 0 - 0.03$ \lsco}
\begin{figure}[t]
\centerline{
\epsfig{figure=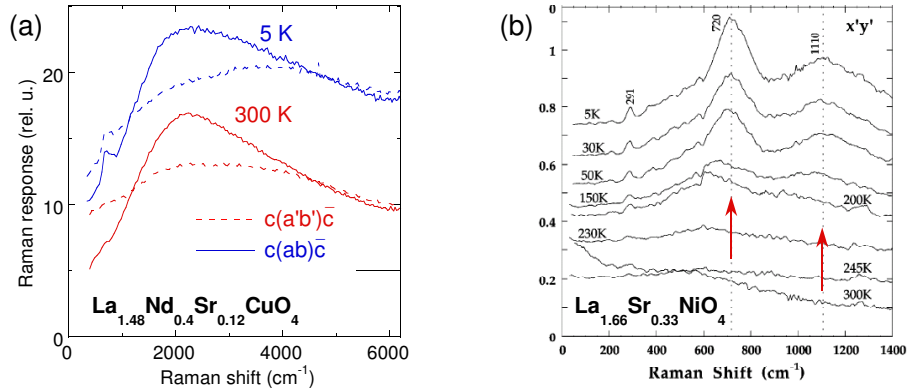,width=120mm}
}
\caption{
(a) Two-magnon Raman scattering in \lns0418 for T~=~300 (red) and 5~K (blue).
The T~=~5~K spectra is vertically off-set.
Dashed (solid) lines are for spectra in $c(a'b'){\bar c}$ and $c(ab){\bar c}$ polarizations respectively.
(b) Two-magnon scattering in the striped phase of La$_{\frac{5}{3}}$Sr$_{\frac{1}{3}}$NiO$_{4}$ from Ref.~\cite{BlumbergPRL98}.
The two red arrows stand for two magnetic bands corresponding to two spin exchange channels, see text.
}
\label{f121}
\end{figure}

Two-magnon (2M) Raman scattering provides an additional way to look at the effects of stripe correlations on magnetic excitations. 
For 2D square lattices the 2M peak is predicted to be seen in the B$_{1g}$ channel, probed by $(ab)$ polarization \cite{FleuryPR68ShastryPRL90}.
Fig.~\ref{f121} shows 2M scattering around 2200~\cm-1 at 300 and 5~K taken with the resonant $\omega_{in}$~=~3.05~eV incident frequency.
As in other tetragonal 2D AF's \cite{SugaiPRB90} we observe the spin pair excitations in the expected scattering geometry.
The $c(a'b'){\bar c}$ polarization shows a featureless background which probably has a contribution from luminescence.

In La$_{2-x}$Sr$_{x}$NiO$_{4}$ there is a clear signature of the effect of stripe ordering on the high energy spin pair excitations: in the undoped case (x~=~0) the 2M Raman band is seen around 1650~\cm-1 \cite{SugaiPRB90}.
At 33\% Sr doping this excitation is not present at that frequency but instead two peaks at lower energies, 720 and 1110~\cm-1 \cite{BlumbergPRL98}, are observed below the magnetic ordering temperature, see Fig.~\ref{f121}b.
In Ref.~\cite{BlumbergPRL98}, assuming an unrenormalized value for the superexchange $J \approx 240$~\cm-1 with respect to the undoped case, it is proposed that these peaks originate from the two spin exchange channels opened due to the stripe order, one of them within and the other one across the antiphase AF domains depicted in Fig.~\ref{f116}.
A more recent neutron scattering study, whose authors are however in favor of a renormalization of the magnetic super-exchange in the stripe phase, is in support of this assignment regarding the higher frequency peak at 1110~\cm-1 ($\approx 2 \cdot 70$~meV) by finding that the upper edge of the spin-wave dispersion branch is around 70~meV \cite{BoothroydPRB03}.
Irrespective of the microscopic origin, the 2M scattering is definitely a good probe for the study of local effects induced by the stripe order.
Comparison with our high energy Raman spectra shown in Fig.~\ref{f121}a shows, as in the case of phonons, that in \lnsco we observe only slight changes from 300 to 10~K emphasizing weak local spin modulations in this compound.

The differences we observe between cuprates and nickelates can be related to the much stronger carrier
self-confinement in the latter \cite{AnisimovPRL92}.
It has also been shown \cite{McQueeneyPRL01} that anomalies in phonon dispersions occur in \lsco at points in the Brillouin zone commensurate with charge ordering wavevectors inferred from neutron scattering studies.
But as discussed, the charge modulation in Nd doped \lsco, where the stripe correlations were shown to be stabilized, is too weak to produce observable changes in the lattice unit cell.
The number of phononic modes we observe can be explained solely in terms of LTO/LTT distortions.
Our data, however, do not contradict the possible existence of charge modulations in the CuO$_{2}$ plane. In fact, the dynamics in the cation-O layers and the magnitude of octahedra tilt disorder affects 
the carrier distribution and our Raman results impose constraints on the magnitude of the charge modulations.
\begin{figure}[t]
\centerline{
\epsfig{figure=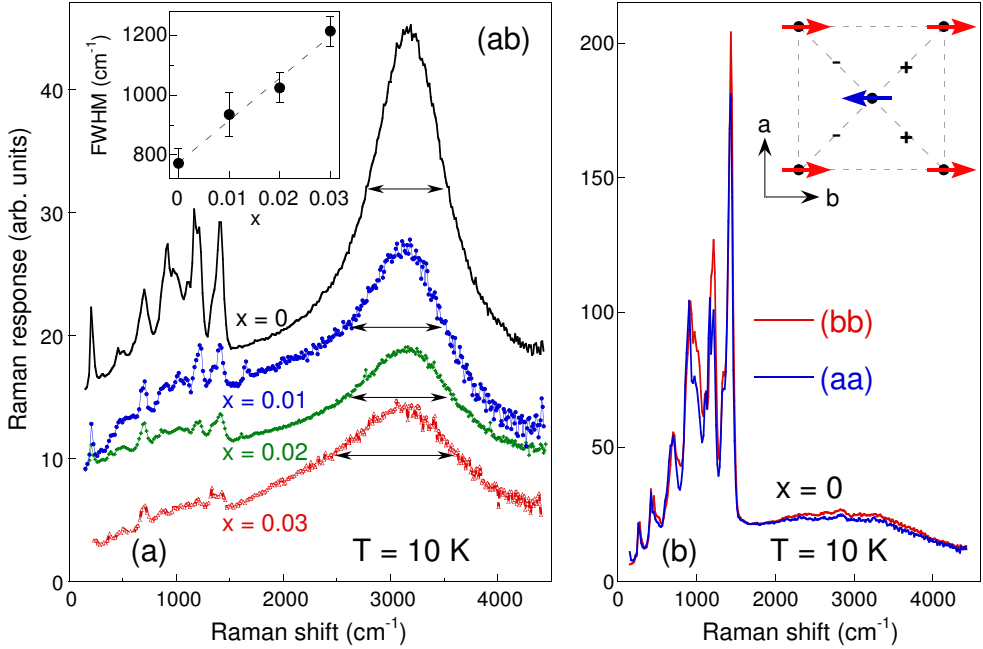,width=130mm}
}
\caption{
T~=~10~K data in \lsco.
(a) The Raman two-magnon excitation around 3000~\cm-1 in $(ab)$ polarization for $x~=~0~-~0.03$.
Spectra are vertically offset.
The sharp features below about 1500~\cm-1 are higher order phonons.
The inset shows the full width at half maximum for the four dopings.
(b) Raman data in a detwinned \lco crystal (see section 1.2 in this chapter) in $(aa)$ and $(bb)$ polarizations.
Note that the strong two-magnon feature is absent in this polarization and the anisotropy of the spectra at high frequencies.
}
\label{f122}
\end{figure}

In the context of sensitive short wavelength magnetic excitations to possible spin/charge modulations in the transition metal oxygen planes, see Fig.~\ref{f121}, and also the strong anisotropy effects found in detwinned \lsco crystals in terms of both phononic and low lying electronic excitations, see Fig.~\ref{f122}, it is interesting to take a look at the evolution with hole concentration of the 2M excitation in lightly doped \lsco crystals as well as to check whether one can find at high frequencies (2000~-~4000~\cm-1) anisotropy effects similar to those observed in Fig.~\ref{f115}.
Our Raman results regarding these problems are shown in Fig.~\ref{f122}.
In panel (a) we show the 2M feature at low temperatures in the B$_{1g}$ channel (probed in $(ab)$ polarization, see the axes notations in Figs.~\ref{f12} and \ref{f117}) for four dopings.

The length of the arrows roughly indicate the evolution of the scattering width with increasing $x$ from 0 to 0.03.
We found quite a sizeable effect in terms of the 2M FWHM.
In the undoped sample the FWHM is around 770~\cm-1 and although the in plane correlation lengths remain large in the lightly doped regime~\cite{KeimerPRB92}, we observe an approximately linear increase of about 60\% with doping from $x = 0$ to 0.03 (inset of Fig.~\ref{f122}b).
One may try to correlate this to the increase in the low energy electronic Raman background seen in Fig.~\ref{f115} in parallel polarization.
We remark that although this enhancement is obvious for $(aa)$ and $(bb)$ scattering geometries, we checked that in $(ab)$ configuration the intensity of the Raman background at T~=~10~K is doping independent.
It is easier to understand that the magnetic spin-wave gap excitations in the long wavelength limit disappear together with the disappearance of long range magnetic order, however, the drastic change in the two-magnon scattering, which in principle requires 'good' AF correlations on a much smaller scale (about 4 lattice constants), is not so straightforward to grasp.
For an explanation one may resort again to the argument that in the CuO$_{2}$ plane, a hole entering the O$2p$ bands is more delocalized and one hole breaks effectively more magnetic bonds than just one between a pair of nearest neighbor spins.

Regarding possible anisotropy effects, Fig.~\ref{f122}b shows that the $(aa)$ and $(bb)$ polarized spectra look very much alike at high frequencies and although there are differences, the second and third order phononic scattering is much less sensitive to the macroscopic orthorhombicity than the one phonon excitations in the 0 to 500~\cm-1 region.
Unfortunately, as predicted by the Fleury and Loudon (see Ref.~\cite{FleuryPR68ShastryPRL90}) and seen in Fig.~\ref{f122}, the strongest 2M scattering is supposed to be seen in B$_{1g}$ tetragonal channel which in not probed in $(aa)$ and $(bb)$ polarizations ($(aa)$ and $(bb)$ configurations probe A$_{1g}$~+~B$_{2g}$ tetragonal symmetries) so we cannot directly probe the effects of the macroscopic lattice spin anisotropy on the strong 2M feature from Fig.~\ref{f122}a.
What we can however say is that if the finite Raman background between 2000 and 4000~\cm-1 has some fully symmetric (A$_{1g}$) magnetic contribution from higher order light scattering Hamiltonian, see Ref.~\cite{SulewskyPRL91}, the influence of the lattice orthorhombicity on that contribution is negligible.

\section{Summary}
Two aspects in connection with the magnetic properties of \lnsco single crystals were discussed in some detail.
One of them was related to long wavelength magnetic excitations in $x~=~0$, 0.01, and 0.03 \lsco detwinned crystals as a function of doping, temperature and magnetic field.
Two magnetic modes were observed within the AF region of the phase diagram.
The one at lower energies was identified with the spin-wave gap induced by the antisymmetric DM interaction and its anisotropic properties in magnetic field could be well explained using a canonical form of the spin Hamiltonian.
A new finding was a magnetic field induced mode whose dynamics allowed us to discover a spin ordered state outside the AF order which was shown to persist in a 9~T field as high as 100~K above the
N\'{e}el temperature T$_{N}$ for x~=~0.01.
We proposed for the field induced magnetic order a state with a net WF moment in the CuO$_{2}$ plane and analyzed the field induced modes in the context of in-plane magnetic anisotropy.
For these single magnon excitations we mapped out the Raman selection rules in magnetic fields and we also found that their temperature dependent spectral weight (in the presence of a constant external magnetic field) was peaked at the N\'{e}el temperature.

The second aspect was related to phononic and magnetic Raman scattering in La$_{2-x-y}$Nd$_{y}$Sr$_{x}$CuO$_{4}$ with three doping concentrations: $x \approx 1/8$, $y =
0$;  $x \approx 1/8$, $y = 0.4$; and $x = 0.01$, $y = 0$.
We observed that around $1/8$ Sr doping and independent of Nd concentration there exists substantial disorder in the tilt pattern of the CuO$_{6}$ octahedra in both the orthorhombic and tetragonal phases which persist down to 10~K and are coupled to bond disorder in the cation layers.
The weak magnitude of existing charge/spin modulations in the Nd doped structure did not allow us to detect specific Raman signatures on lattice dynamics or two-magnon scattering around 2200~\cm-1.

It is possible that the discovery of weak charge modulations in the hole doped 2D CuO$_{2}$ planes characteristic of high T$_{c}$ materials is just a matter of time.
The problem of doped Mott-Hubbard insulators seems however to be one in which numerous possible ground states are allowed and the supremacy of any one of them could require really fine tuning of microscopic parameters.
In this respect, even if such a charge density modulation were observed, the question whether it helps understanding the mechanism of superconductivity or not would still need to be answered. 

{\bf Acknowledgments --}
We acknowledge discussions and collaborations with B. S. Dennis, M. V. Klein and A. N. Lavrov. 


\end{document}